\documentclass[aps,prl,reprint,superscriptaddress]{revtex4-1}

\usepackage{amsfonts}
\usepackage{amsmath}
\usepackage{graphicx}
\usepackage[utf8]{inputenc}
\usepackage{amssymb}
\usepackage{times}
\usepackage{color}

\begin{document}

\title{From three-photon GHZ states to ballistic universal quantum computation}
\author{Mercedes Gimeno-Segovia}
\affiliation{Department of Physics, Imperial College London, London SW7 2AZ, United Kingdom}
\author{Pete Shadbolt}
\affiliation{Department of Physics, Imperial College London, London SW7 2AZ, United Kingdom}
\author{Dan E. Browne}
\affiliation{Department of Physics and Astronomy, University College London, London WC1E 6BT, United Kingdom}
\author{Terry Rudolph}
\affiliation{Department of Physics, Imperial College London, London SW7 2AZ, United Kingdom}
\date{\today}

\begin{abstract}
{Single photons, manipulated using integrated linear optics, constitute a promising platform for universal quantum computation. A series of increasingly efficient proposals have shown linear-optical quantum computing to be formally scalable. However, existing schemes typically require extensive adaptive switching, which is experimentally challenging and noisy, thousands of photon sources per renormalized qubit, and/or large quantum memories for repeat-until-success strategies. Our work overcomes all these problems. We present a scheme to construct a cluster state universal for quantum computation, which uses no adaptive switching, no large memories, and which is at least an order of magnitude more resource-efficient than previous passive schemes.  Unlike previous proposals, it is constructed entirely from loss-detecting gates and offers a robustness to photon loss. Even without the use of an active loss-tolerant encoding, our scheme naturally tolerates a total loss rate  $\sim 1.6$\% in the photons detected in the gates.} This scheme uses only 3-GHZ states as a resource, together with a passive linear-optical network. We fully describe and model the iterative process of cluster generation, including photon loss and gate failure. This demonstrates that building a linear optical quantum computer need be less challenging than previously thought.
\end{abstract}
\maketitle

In 2001, Knill, Laflamme and Milburn \cite{Knill1998} showed that scalable quantum computation was possible using only linear optical elements — without the need for deterministic two-photon interactions. However, their proposal was more a proof of principle than a feasible construction as the scheme required tens of thousands of optical elements to acquire gates with a high probability of success. Since then, several proposals have developed the idea of a linear optical quantum computer (LOQC), including Nielsen's proposal \cite{Nielsen2004} of combining linear optics with cluster states, Browne and Rudolph's fusion mechanisms \cite{Browne2005} to efficiently create optical cluster states and Kieling's et al proposal \cite{Kieling2007}  of building an imperfect cluster that can be renormalized using ideas of percolation theory.  While alternative schemes for LOQC \cite{Kok2007} using parity state encoding \cite{Hayes2010} or small amplitude coherent states \cite{Lund2008} have been proposed, we do not address these approaches in this manuscript.

Recent demonstrations \cite{carolan2014,silverstone2014,spagnolo2014,cai2014,barz2014} have made significant progress towards experimental linear-optical quantum computing. In particular, the use of integrated photonics to implement large-scale, complex interferometers on a chip shows great promise. However, active feed-forward remains challenging, it requires fast switching which is a dominant source of photon loss and has not yet been experimentally demonstrated in an integrated device. 

Of previous approaches to linear optical quantum computing, only Kieling et al's proposal \cite{Kieling2007} is \textit{ballistic} - meaning that active switching is not required for the process of cluster state generation. It is thus the most suitable previous approach to LOQC in an integrated setting. It has a number of shortcomings, however. Firstly, it requires 4 or 5-photon entangled states as input---costly and difficult to generate in a (near)-deterministic manner. Secondly, it is not constructed from loss-tolerant components, photon loss during the process will lead to the generation of an undesired state.

In this Letter, we adapt new advances in Bell state measurement \cite{Grice2011,Ewert2014} to the ballistic cluster state generation scheme to provide a new approach to scalable ballistic LOQC with significant advances on Kieling et al's approach. Off-line resources are reduced to 3-photon entangled states, while all gates are loss-detecting. The scheme has an in-built robustness to loss and will succeed, without additional loss-encoding, even if $>1\%$ of the photons entering the gates are lost. Deterministic $n$-qubit entangled state generation becomes increasingly experimentally challenging with $n$ \footnote{The best known theoretical GHZ building strategy from Bell Pairs has a success probability $p_{succ}=\left(\frac{1}{2}\right)^{\lfloor\frac{n-1}{2}\rfloor }\left(\frac{3}{4}\right)^{\lceil\frac{n-1}{2}\rceil }$. Gimeno-Segovia et al. (in preparation)} , and the reduction to resource states of only 3 photons is thus a significant improvement.  For a fair comparison of our scheme against previous proposals \cite{Kieling2007} we count the number of Bell pairs needed to build the initial entangled states for both cases. As the construction of these initial states is probabilistic, we assume a multiplexing stage in order to achieve deterministic resource states, which then enables us to count the total number of Bell pairs used in each strategy. The full resource comparison, demonstrating at least an order of magnitude reduction in resources, is presented in the supplementary material.

\begin{figure}[h]
\centering
\includegraphics[width=\linewidth]{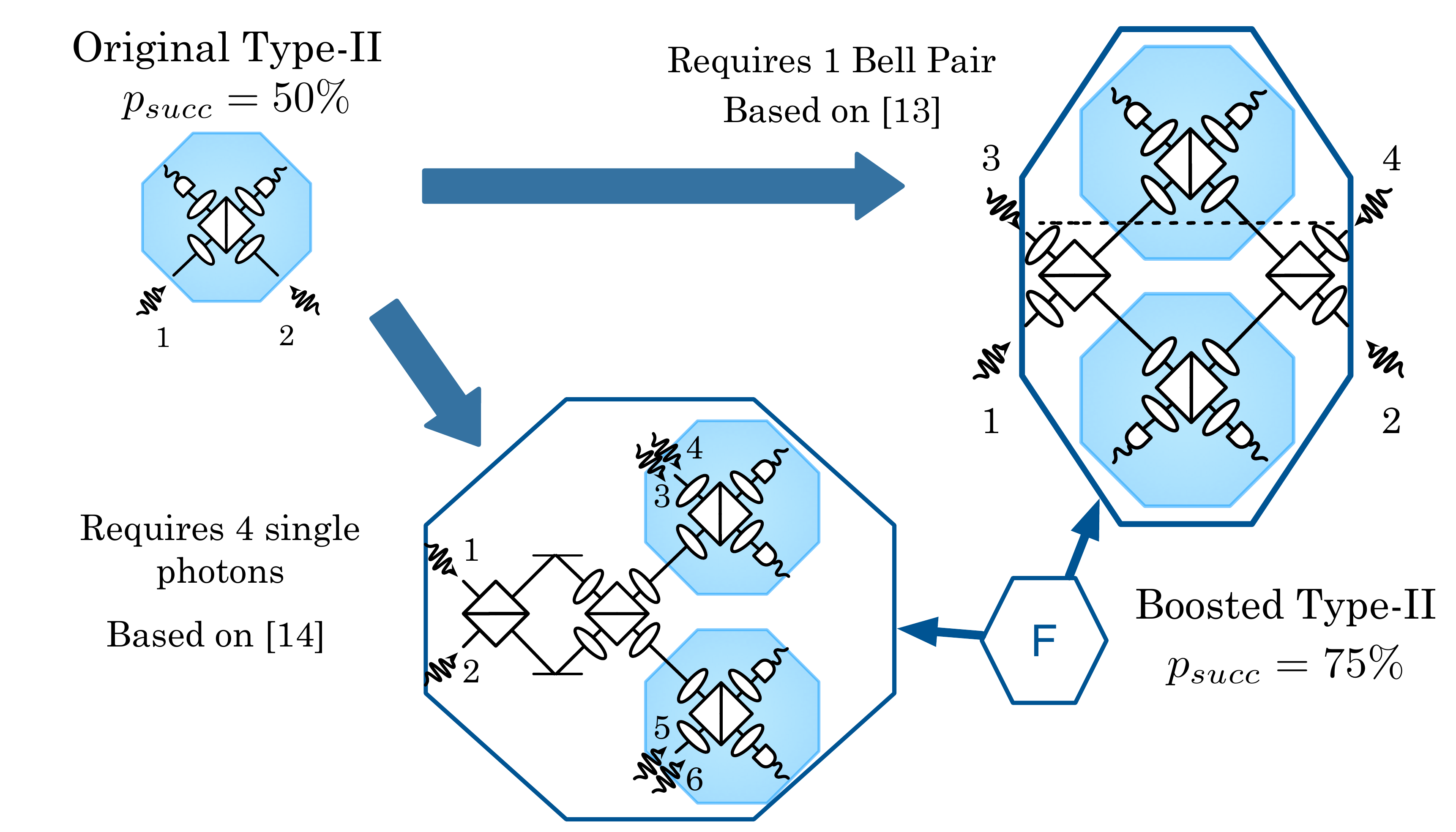}
\caption{(Color online) Boosted Type-II fusion gate. Photons 1 and 2 represent the photons on which the gate is applied, the rest are ancillary photons. The implementation based on \cite{Grice2011} requires a pair of maximally entangled photons, while the implementation based on \cite{Ewert2014} requires 4 single photons. The boosted gates have the exact same success and failure outcomes as the original Type-II but with a higher success probability. Note that all photons are measured.  Here and in subsequent figures we will represent the boosted fusion by the hexagon marked 'F'.}
\label{fig:boosted}
\end{figure}

The basic building block of our scheme is Browne and Rudolph's Type-II fusion gate, which can be used to connect small cluster state fragments into a large cluster state for measurement-based quantum computing. This gate is equivalent to a Bell state measurement (BSM) in a rotated basis. In linear optics, BSMs cannot be achieved deterministically. For a long time, the highest known probability of success for a linear optical BSM was 50\% \cite{calsamiglia2001maximum}, but recent breakthroughs have shown that this can be improved to 75\% by incorporating ancillary resources - such as Bell Pairs \cite{Grice2011} or single photons \cite{Ewert2014}. 
We adapt these schemes to give a Type-II fusion gate with the same enhanced probability. The advantage of using Type-II fusion instead of Type-I as in previous proposals\cite{Kieling2007}, is that this gate detects any lost photons and therefore does not introduce logical errors \cite{Varnava2008}.



The phenomenon of percolation has been long studied \cite{Stauffer1994} in classical statistical mechanics as a prototype phase transition on graphs that have lost some of their bonds and/or sites due to a randomized process with a probability $1-p$. When $p$ is above the percolation threshold, there exists at least one spanning path from one side of the lattice to the other.
 In the context of one-way LOQC, the percolation graph will define a cluster state, whose bonds/sites are effectively removed due to failure of probabilistic entangling gates together with photon loss. The percolation threshold marks a phase transition in the computational power of the resource state generated \cite{Kieling2007a}, which distinguishes the states that can be used for universal quantum computation from those which cannot. 

Here, we exploit the 75\% success probability of the boosted Type-II gate, to develop a new percolation approach in which 3-photon cluster states are fused together to form a lattice. The underlying graph we choose is the diamond lattice, as it has the lowest vertex degree of all 3D lattices and yet it shows good percolation properties in comparison with 2D lattices with the same correlation number per site \cite{Tarasevich1999}.  As it will be shown in figures \ref{fig:LON} and \ref{fig:percolated}, in our scheme, failure of a gate produces correlated bond losses as well as the appearance of bonds that do not belong to the diamond lattice form. This is very different from the uncorrelated bond loss model which is usually studied in statistical mechanics, and therefore we cannot employ existing analytic or numerical results.


\begin{figure}[hbt]
    \begin{center}
    \includegraphics[width=\linewidth]{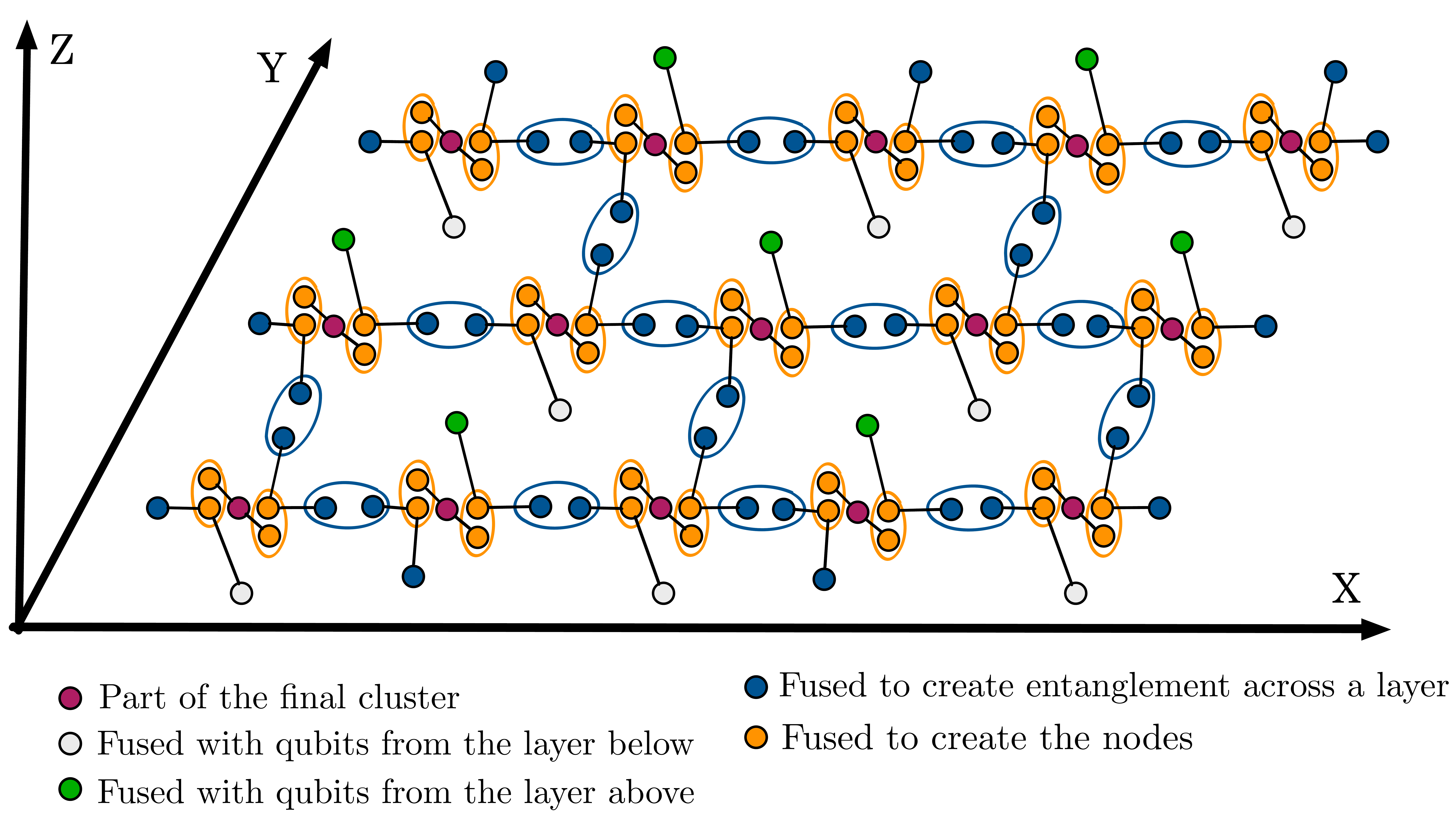}
    \caption{(Color online) Full Layout of a layer of the diamond graph using 3-photon GHZ states as input. The legend at the bottom of the figure shows the role of each photon. There are two types of rotated fusion Type-II gate used, their effect on the GHZ states is described in figures \ref{fig:LON} and \ref{fig:fusingMicroclusters}.}
    \label{fig:FullLayout}
    \end{center}
\end{figure}

The internal structure of a diamond lattice can equivalently be seen as a “brickwork” in three dimensions (Fig. \ref{fig:FullLayout}) . This picture is useful when arranging the microclusters prior to fusion, as all bonds then lie in one of three orthogonal directions. The diamond lattice is formally isotropic, however its brickwork depiction is not, there is a greater average connectivity in the X direction and thus a preferred direction for percolation. The process by which the lattice is generated (figs \ref{fig:LON} and \ref{fig:fusingMicroclusters}) is optimized to take advantage of this anisotropy.

\begin{figure}[hbt]
    \begin{center}
    \includegraphics[width=\linewidth]{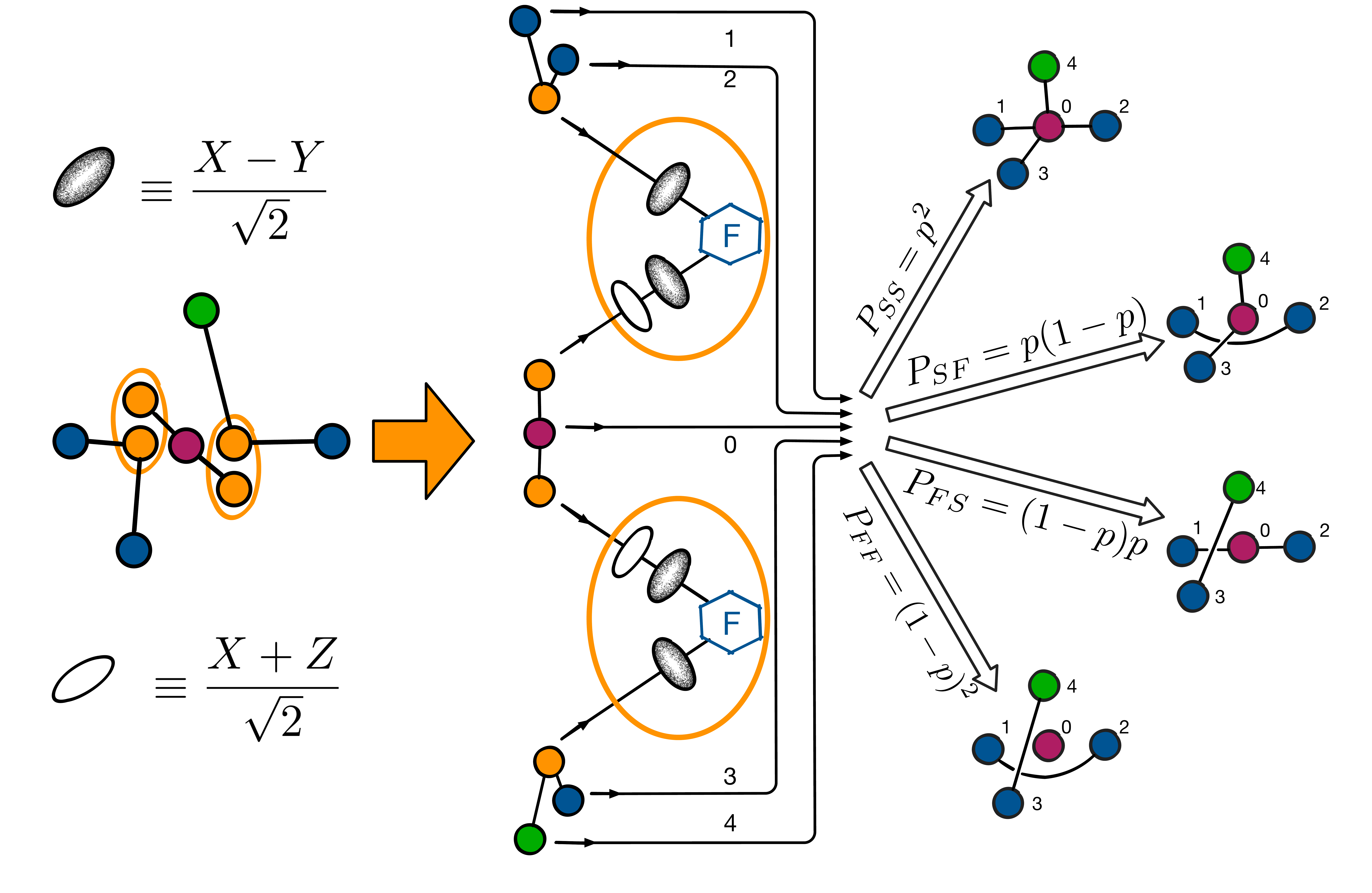}
    \caption{(Color online) Probabilistic creation of star microclusters.}
    \label{fig:LON}
    \end{center}
\end{figure}

In figure \ref{fig:FullLayout}  we can see how the GHZ states are arranged to create the brickwork structure. For each site in the final lattice, we use three 3-GHZ states to create a five-qubit microcluster. Each microcluster is created by performing two rotated Type-II fusion gates \cite{Browne2005}, as described in figure \ref{fig:LON} . The 5-star microcluster will be created when both fusions succeed, however in the case of failure, the outcomes will still create connectivity in the lattice, contributing still to the percolation of the whole lattice. In the case where we have formed a 5 qubit star graph state, all the qubits in the exterior are equivalent, however in the cases where failures have happened, the way in which we arrange those external qubits affects the connectivity of the lattice. We have shown in figure \ref{fig:LON} the arrangement that is most suitable for our scheme and that allows us to obtain the lowest percolation threshold.

\begin{figure}[hbt]
 \begin{center}
\includegraphics[width=\linewidth]{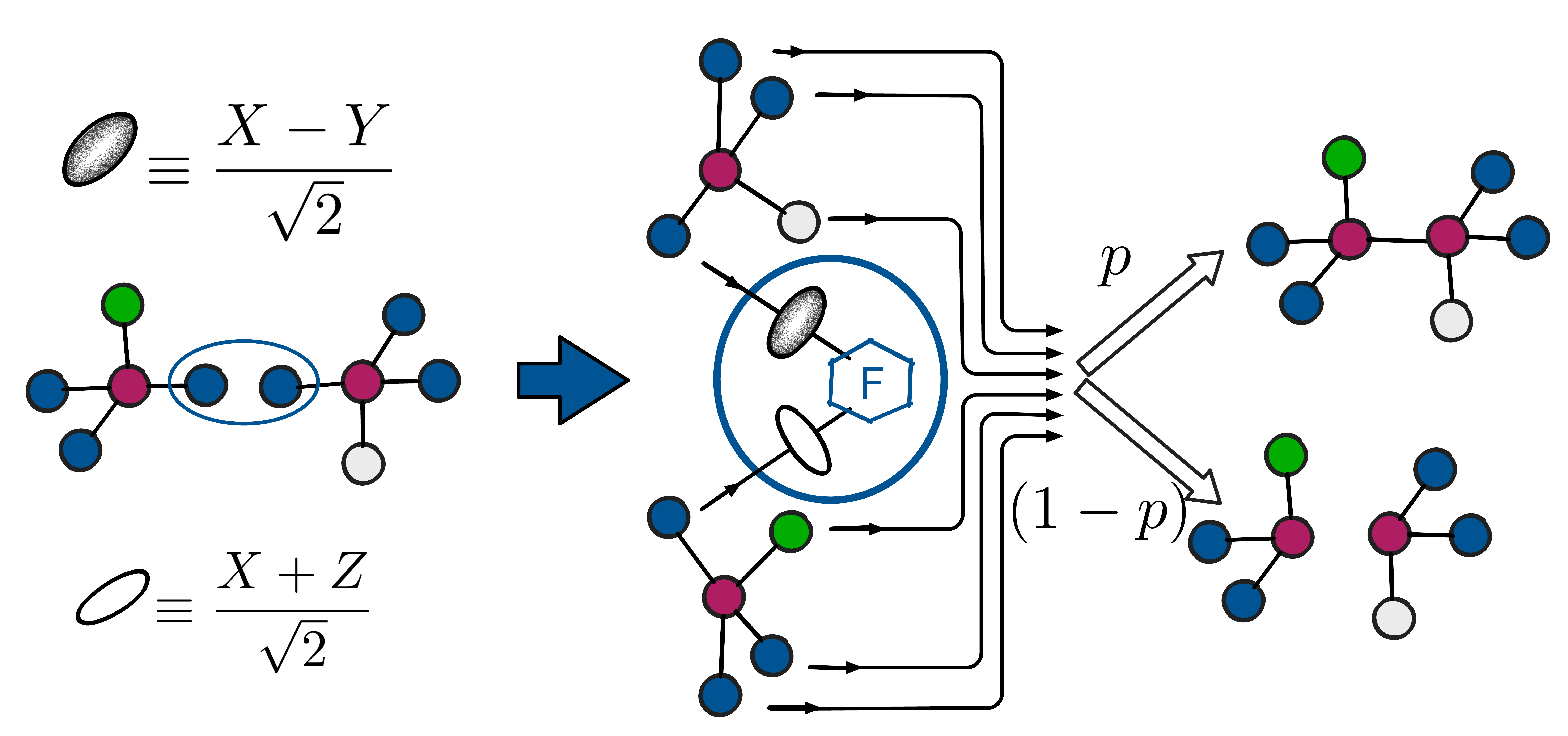}
\end{center}
\caption{(Color online) Fusion of 5-qubit microcluster to form the final lattice}
 \label{fig:fusingMicroclusters}
\end{figure}

To assess the percolation properties of the lattice, we use a Monte Carlo simulation.
In each independent run, our simulation builds the lattice sequentially, modelling the action of the success and failure of the fusion gates and attempts to find a percolation path. In doing so, we achieve a more realistic picture compared to the simpler alternative of deleting nodes from a perfectly formed lattice.  This approach also allows us to observe the information which will ultimately be fed to a classical percolation algorithm. For each set of parameters, the simulation is run $10^4$ times to ensure that statistical error in the data is $\lesssim 1\%$.

\begin{figure}[hbt]
    \begin{center}
    \includegraphics[width=\linewidth]{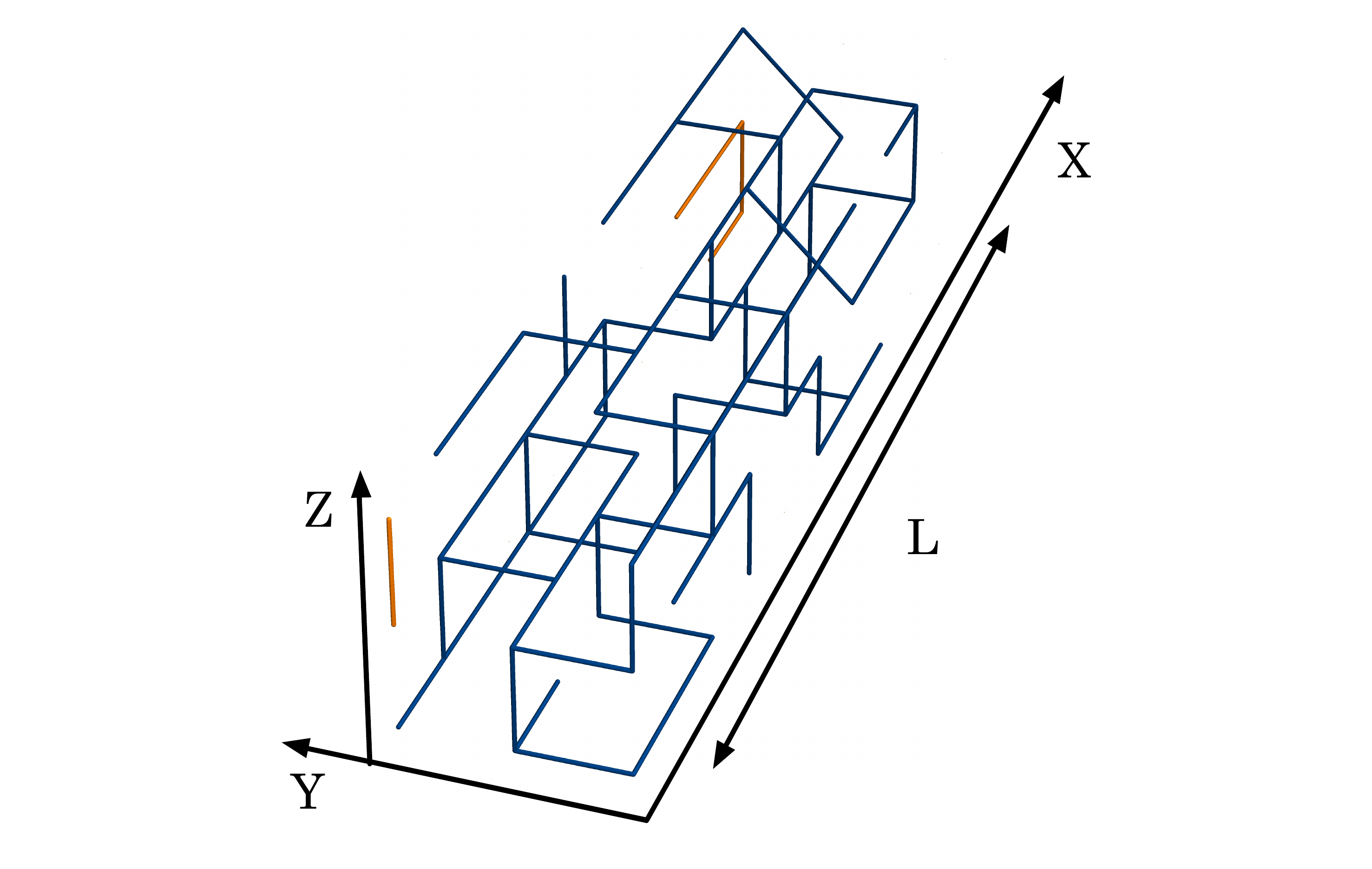}
    \caption{(Color online) Instance of the percolated cluster (10x3x3), highlighted in blue is the spanning cluster. In addition to the orthogonal bonds which are expected in the canonical brickwork lattice, we see some diagonal bonds — these are the result of failed fusions during the creation of microclusters. }
    \label{fig:percolated}
    \end{center}
\end{figure}

 In figure \ref{fig:percolated} we present an instance of the lattice, where we can see why this lattice is not the typical percolated diamond lattice. 
  The failures of some of the fusion gates produce correlated bond losses together with the appearance of new diagonal bonds that can be seen in the figure. It must be noted that the presence or absence of the bonds will be known from the pattern of successes and failures of the fusion gates. Thus in any experimental set up, the structure of the percolated lattice could be inferred by a simple classical algorithm.

Let us define $\Pi\left(p,L\right)$ as the probability that a lattice of linear dimension $L$ percolates when built with fusion gates that succeed with probability $p$. The percolation threshold can be calculated from finite size lattices by finding the crossing point of the function $\Pi(p,L_i)$ for different values of $L_i$ { (a justification for this procedure can be found in the supplementary material)}.
\begin{figure}[hbt]
    \begin{center}
    \includegraphics[width=\linewidth]{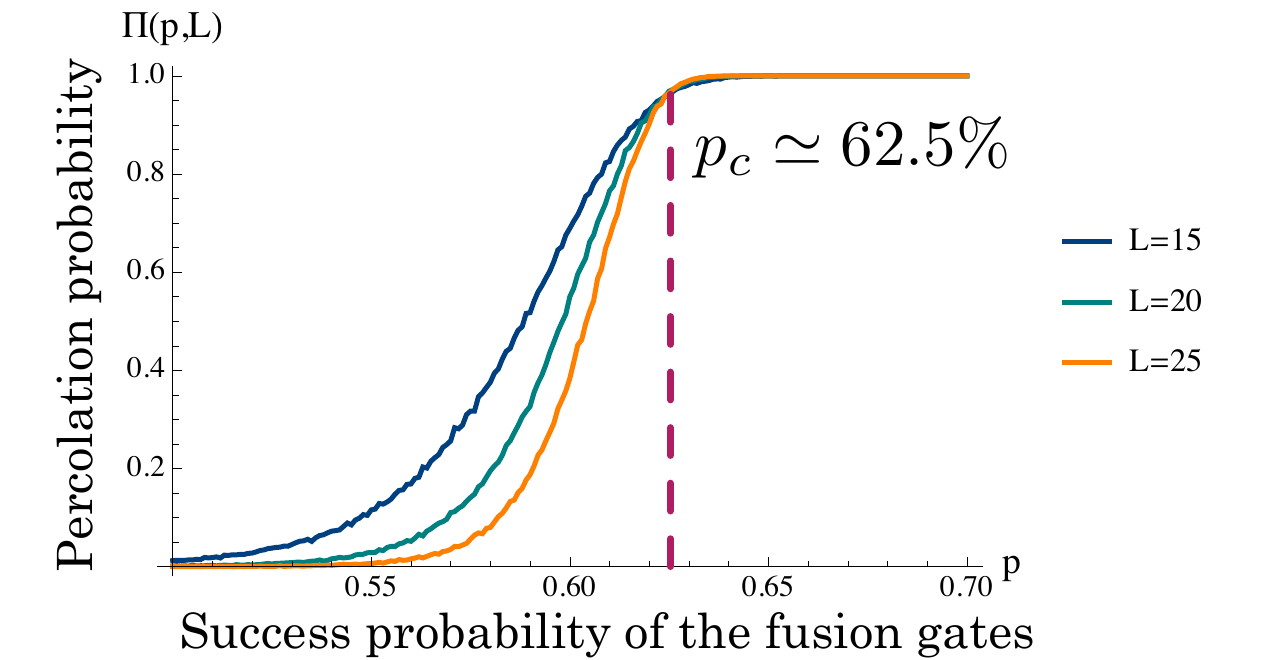}
    \caption{(Color online) Results for simulations on a bulk of cluster of L= 15, 20, 25. Each cluster contain $L^3$ sites and has been generated from $3\cdot L^3$ GHZ states.}
    \label{fig:threshold}
    \end{center}
\end{figure}

We perform the simulation by generating instances of the lattice with fusion success probability $p$. In figure \ref{fig:threshold} we have represented the results for lattices of different linear dimension and find the value for the percolation threshold, which is estimated to be $p_c \simeq 0.625$. We conclude that lattices built according to our scheme, using boosted fusion gates with success probability of $75\%$, are well above the percolation threshold --- and are therefore universal for quantum computing.


\emph{A single qubit channel:} In traditional MBQC, a single qubit is replaced by a linear cluster. When two-qubit operations are required, a bond (gate) is created between two linear clusters (qubits). In a paradigm where the creation of entanglement between qubits is probabilistic (such as in LOQC), a three-dimensional piece of cluster state can be used to implement a single functional qubit.  If there exists a spanning path through the cluster, information can flow through the channel, allowing the computation to progress. We can then calculate how many operations we can perform on this single qubit.
 
The cluster channel is parametrized by a fixed cross section (width and height) and variable length, which 
corresponds to the computational depth. The cross section of this cluster is directly related to its percolation properties ---  a larger cross section gives a higher percolation probability. Given a desired length, we must choose a cross section in order to have a percolation probability higher than some desired probability of success. In figure \ref{fig:long} we show the percolation probability for different cross sections, as a function of the length. We have chosen square cross sections because in preliminary simulations this geometry performed better than rectangular shaped cross sections.
 
 \begin{figure}[hbt]
    \begin{center}
\includegraphics[width=\linewidth]{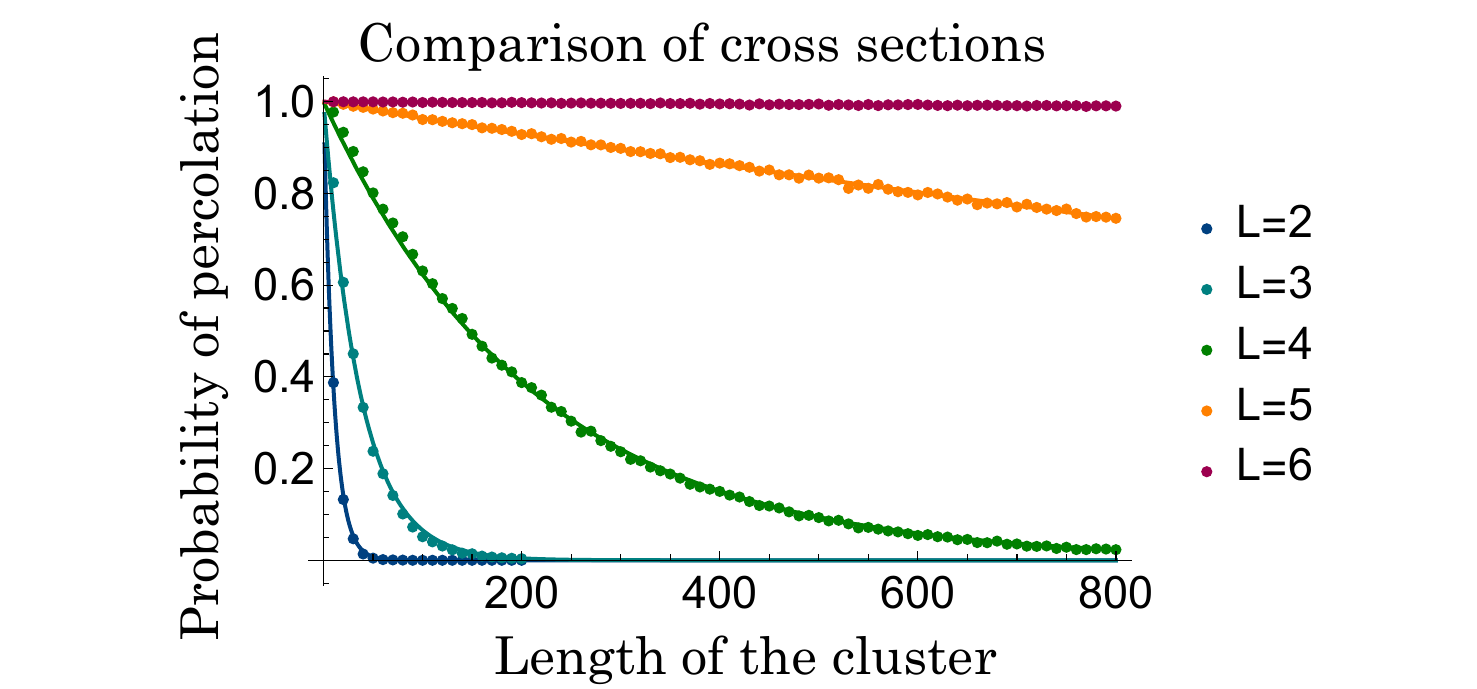}
    \caption{(Color online) Percolation probabilities as a function of the length, for lattices of square cross section $L^2$. The length of the cluster correlates to the computational depth of the lattice. The exponential decay shown has a decay constant which depends quadratically on $L$.}
    \label{fig:long}
    \end{center}
\end{figure}
 
As we can see from figure \ref{fig:long}, for a cross section of 6 $\times$ 6 qubits, we can make the cluster very deep. Because of computational constraints, simulating large clusters is very challenging. We fit an exponential decay function to the data, obtaining an estimated variance of $10^{-7}$. From this fit we extrapolate that for $L=6$ we have more than 90\% probability of percolation in clusters up to 9000 qubits in length.

\begin{figure}[hbt]
    \begin{center}
  \includegraphics[width=0.85\linewidth]{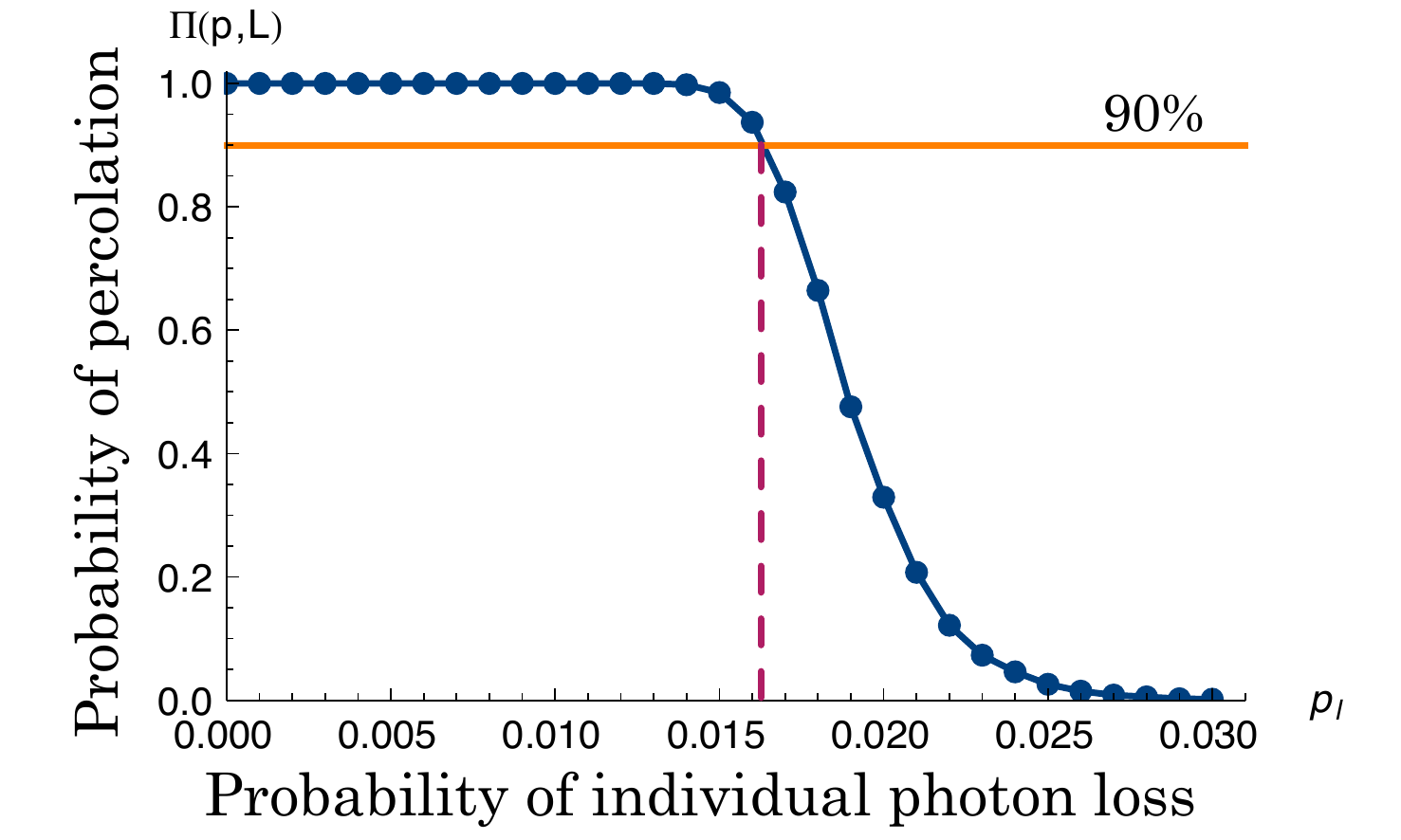}
  \caption{(Color online) Loss tolerance (blue) for a cubic lattice of linear dimension $L=25$.}
  \label{fig:loss}
    \end{center}
\end{figure}

A question that naturally arises in large-scale schemes for LOQC is tolerance to photon loss. This scheme has been designed with loss robustness from the outset. The Type-II boosted fusion gates can detect all losses that happen in the photons incident in the fusion gates. Our scheme is operating above the percolation threshold for the lattice, and this headroom leads to a natural loss tolerance. The incoherence induced in the state by a loss error can be fixed by measuring neighbours of lost qubits in the Z basis, thus cutting bonds from our graph.  We have simulated the building of the lattice where each photon has probability $p_l$ of being lost, and when a loss is detected, we measure all neighbours of the lost qubits in the Z basis to cut it out. In figure \ref{fig:loss}, we can see the loss tolerance of a cubic lattice of $L=25$ in blue, in orange we have highlighted the constant success probability of 90\% for comparison. The success probability of the fusion gates used has been taken to be 75\%. As we can see, the probability or having a spanning path is larger than 90\% for loss rates of up to 1.6\%.

 We want to stress that this is a natural loss tolerance of the system. Previous proposals \cite{Kieling2007} have given thresholds for heralded loss, where the location of all loss errors in the final lattice is known. Heralded loss is not experimentally justified in LOQC and only serves as an upper bound for loss tolerance. In order to compare our scheme with previous work we have performed the same kind of heralded loss simulations and found that in this scenario we could tolerate loss rates up to 15\%, which is an improvement { of 5\% }on the numerical results reported in \cite{Kieling2007}. The improvement over previous proposals \cite{Kieling2007} is not only on the overall robustness of the construction, which is indicated by the 5\% improvement on the heralded loss tolerance, but also the reduction of the amount of resources needed by at least an order of magnitude.

We have presented a ballistic scheme for the construction of a linear optical cluster state that is universal for MBQC. While we have not explicitly included error-correcting codes to provide robustness to loss and errors in the photons in the final computational cluster state, the universality of the cluster state implies a number of ways forward, incorporating tree-clusters \cite{treecluster} or the surface code \cite{surface,topological} as loss-error and general-error correcting codes. Raussendorf's 3D cluster encoded surface code \cite{raussendorf2006}, in particular, seems well suited to ballistic generation. 

To implement this scheme with only 3-photon GHZ as resources we have proposed a boosted fusion mechanism based on \cite{Grice2011} and \cite{Ewert2014} that works with 75\% probability, which is well above the percolation threshold ($p_c=62.5\%$) of this lattice. We have shown the robustness of the scheme in the presence of small amounts of photon loss (up to 1.6\%) and its favourable resource scaling.  Even though this scheme was devised with linear optics in mind, it applies for any physical system with probabilistic gates, and if that probability is higher than 75\% it might be conceivable that the resources needed could be reduced even further. 

For this scheme to be implemented experimentally, it would need a near-deterministic 3-photon GHZ source.  It is not yet known what the optimal way of producing these photonic states is, options range from multiplexing a linear optical circuit such as that proposed in \cite{Varnava2008}, using a similar scheme to the multiplexed single photon source such as \cite{Bonneau2014}, to producing a 3-photon linear cluster (local Clifford equivalent to a GHZ) with a quantum dot \cite{Lindner2009}. As any linear optical fully loss detecting gate must necessarily measure all photons incident on it, the 3-photon GHZ is the minimal resource for a loss-detecting BSM-based ballistic scheme.

Ballistic generation of cluster states for MBQC remains the most attractive approach to scalable linear optical quantum computing. By developing a loss-tolerant and significantly more resource efficient scheme, we have shown that new theoretical ideas continue to ameliorate the technical challenges of building a scalable linear optical quantum computer.

The authors would like to thank Hussain Zaidi, Aida Moreno-Moral, Martik Aghajanian, Chris Dawson and Gabriel Mendoza for helpful discussions. TR and PS supported by the Vienna Science and Technology Fund (WWTF, grant ICT 12-041) and the Army Research Office (ARO) grant No. W911NF-14-1-0133. MGS supported by EPSRC. The numerical simulations were possible thanks to the High Performance Cluster of Imperial College. We would like to draw the reader's attention to the concurrent work in \cite{Zaidi2014} which proposes an alternative approach to this problem.

\clearpage
\widetext
\begin{center}
\textbf{\large From three-photon GHZ states to ballistic universal quantum computation --- Supplementary material}
\end{center}

\section{Calculating the percolation threshold from finite size lattices}
Let us define $\Pi\left(p,L\right)$ as the probability that a lattice of linear dimension $L$ percolates when built with fusion gates that succeed with probability $p$. In the case of the infinite lattice we would have that $\Pi=0$ if $p<p_c$ and $\Pi=1$ if $p>p_c$, however in the case of a finite lattice $\Pi(p)$ for a set $L$ will be a smooth function instead of the Heaviside step function due to finite-size corrections. 

To find the percolation threshold without making any assumptions about the functional form of $\Pi(p,L)$ we use known results about the critical point in the context of renormalisation. The basic idea of renormalisation is the self-similarity of the lattice at the critical point (percolation threshold) \cite{Stauffer1994}. The correlation length , $\xi$ ,can be defined as the typical cluster diameter, it diverges at the percolation threshold as that is the point where the infinite cluster first appears \cite{Grimmett1997} and it grows monotonically with the occupancy $p$. What this means is that the size of any cluster at the percolation threshold is much smaller than the correlation length at the percolation threshold (which is infinite) and therefore all the clusters are similar to each other in an average sense.

When performing renormalisation on a lattice, we replace a cell of the lattice (that involves many sites) by a supersite, provided that the linear dimension of the cell $b$ is much smaller than the correlation length of the lattice $\xi$. At the percolation threshold, because of the self-similarity of large lattices, the properties of a renormalised lattice will be the same as the original lattice and therefore we will have $\Pi(p_c,L)=\Pi(p_c,L/b)$. That is to say that the value of $\Pi(p_c,L)$ does not depend on the renormalisation parameter and must therefore have the same value for all lattices of different size but with the same shape and dimension. We can thus conclude that to calculate the percolation threshold when we only have access to data in finite lattices, we should obtain values of $\Pi(p,L)$ for different  $p$ and $L_i$ and we find the threshold by estimating were the functions of $\Pi(p,L_i)$ intersect.

\section{Renormalisation of the lattice and scaling of resources}
\begin{figure}[hbt]
\begin{center}
  \includegraphics[width=0.75\linewidth]{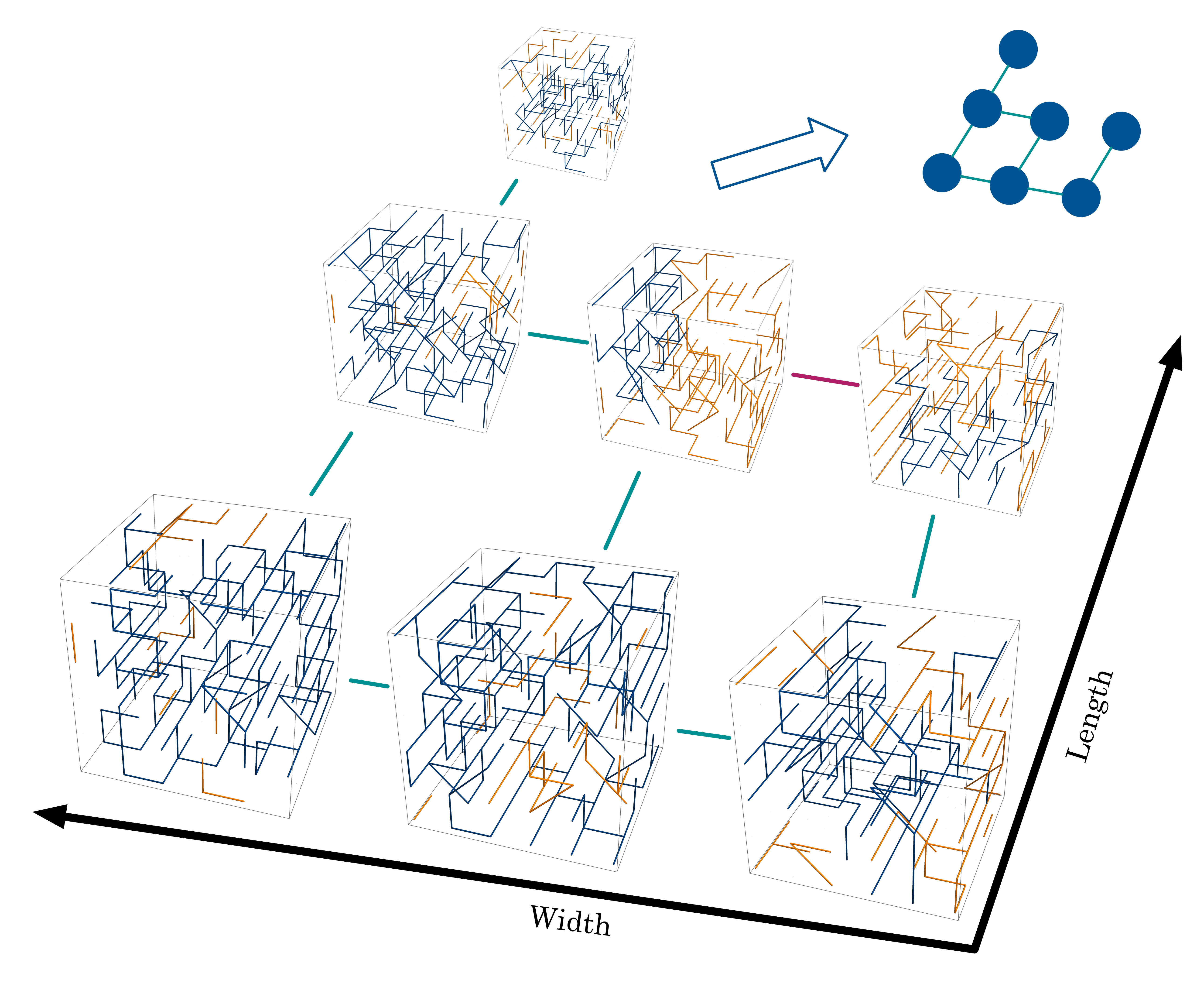}
  \caption{View of the lattice in terms of the renormalised qubits, which are cubic pieces of the lattice. }
  \label{fig:renormalisation}
  \end{center}
\end{figure}

In this section we will explain the renormalisation of the lattice and assess the scaling, assuming that we have GHZ and Bell pairs on demand. In figure \ref{fig:renormalisation} we show the procedure of renormalizing the lattice: we take cubic pieces of the lattice and treat each of them as a renormalised qubit. When fusing the qubits that lie on the sides of these cubes, we are applying CZ gates in the renormalised lattice.

 In figure \ref{fig:renormalisation} we can see a realistic example of the renormalisation procedure, in which cubic pieces of the lattice become the new renormalised qubits. Within these renormalised qubits, the part of the lattice highlighted in blue shows the spanning cluster while the orange highlights the disconnected parts of the cluster. As mentioned in the main body of the article, the lattice percolates better along the length in comparison with the other directions, which means that sometimes the renormalised qubit will be able to connect to other renormalised qubits along the length but not in other directions. When this is the case, we won't be able to perform a CZ gate between the renormalised qubits (shown in figure \ref{fig:renormalisation} in red). These CZ gates that connect qubits across the width correspond to gates between logical qubits, which are more flexible as we can delay them or reconfigure the circuit slightly, therefore not posing a significant problem for the scheme.

\begin{figure}[hbt]
\begin{center}
  \includegraphics[width=0.5\linewidth]{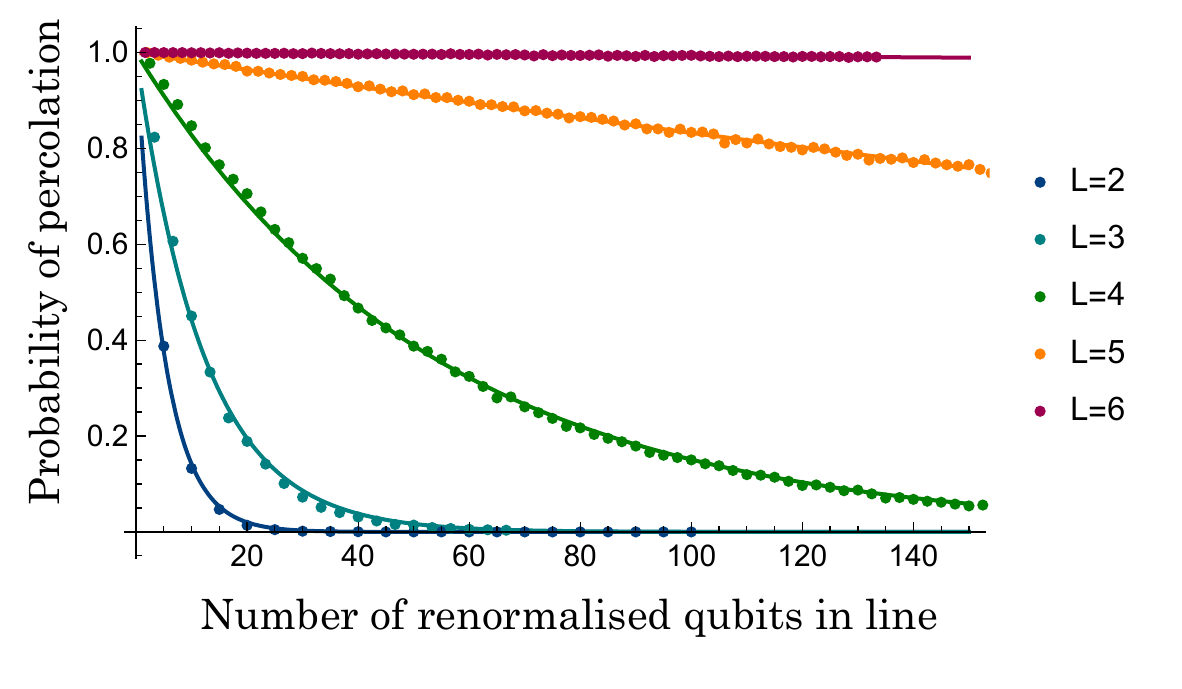}
  \caption{Probability of having a percolating path and therefore information flow, after fusing a number of renormalised qubits in line. The probability shows an exponential decay with the number of qubits fused, however for $L\geq 6$ the decay seems negligible when fusing $O(100)$ renormalised qubits.}
  \label{fig:long}
  \end{center}
\end{figure}

Lets assume that our renormalised qubit is a cubic section of the lattice containing $L^3$ physical qubits. We renormalise figure \ref{fig:long} to show what is the probability of percolation when we fuse a certain number of these renormalised qubits. When we extrapolate the results for $L=6$ to see how many we can fuse before the probability drops below 90\%, we find that we can have a computational depth of 1500 before this occurs. For $L>6$ the computational depth that can be achieved will be much higher.
   
\begin{figure}[hbt]
\begin{center}
  \includegraphics[width=0.5\linewidth]{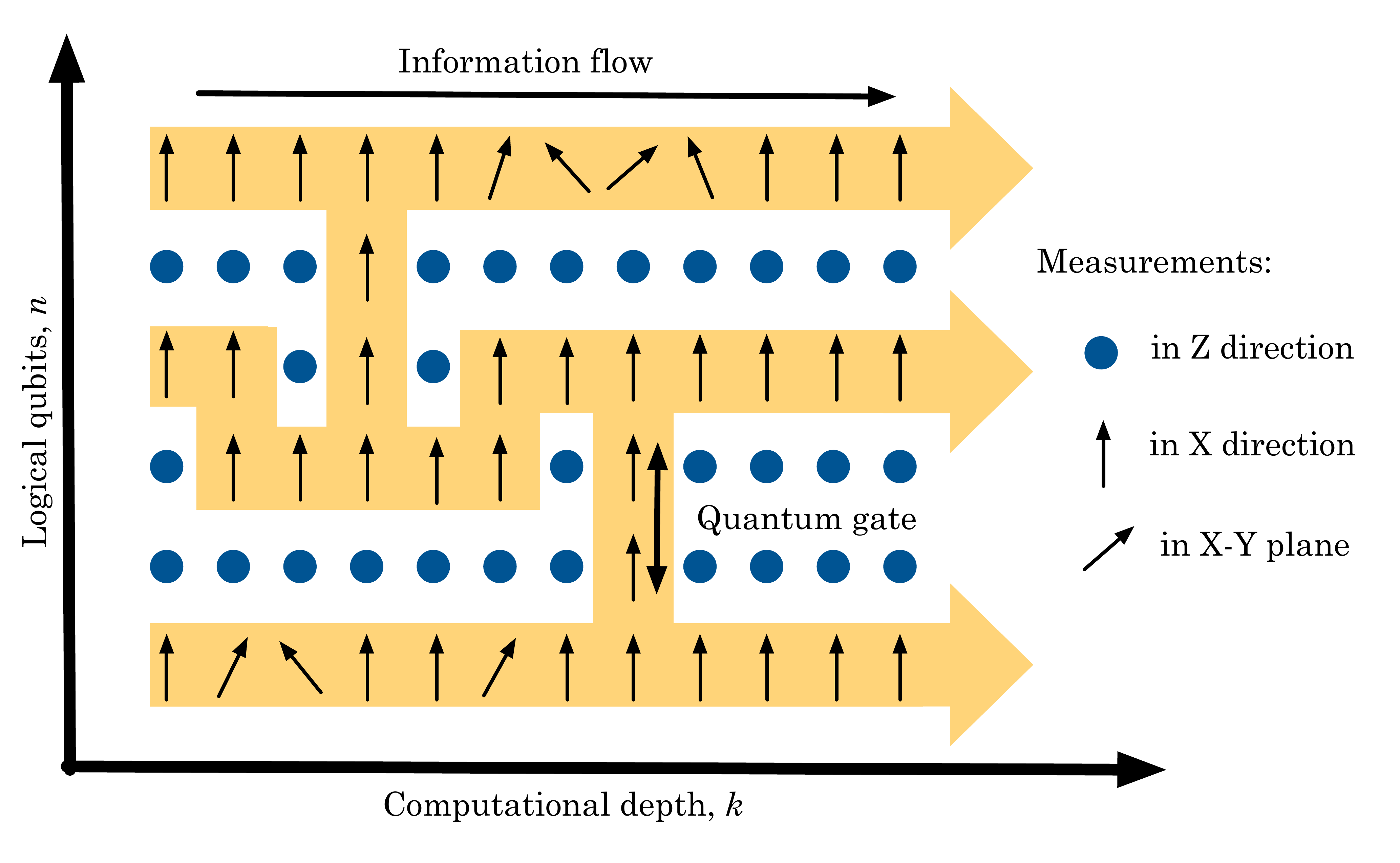}
  \caption{Notation as it relates to a one-way quantum computation \cite{mbqc}, copyright (2001) by the APS . Horizontal arrows indicate the information flow and can be thought as each of them being a logical qubit. Vertical  lines represent quantum gates performed between the logical qubits. The arrows show the direction in which the qubit is to be measured in.}
  \label{fig:oneway}
  \end{center}
\end{figure}

We want a code with computational depth of $k$ (by which we mean the number of measurements we will want to perform in each logical qubit; in the MBQC model, this corresponds to the number of qubits on one line of code). We also want to have $n$ logical qubits (see figure \ref{fig:oneway} where notation is more clearly indicated).
For these variables, the values of the quantities involved and resources needed is:
\begin{itemize}
    \item Number of renormalised qubits: $(n\cdot k)$
    \item Total number of lattice sites: $(n\cdot k)\  L^3$
    \item Number of 3-photon GHZ needed: $3(n\cdot k )\ L^3$
    \item Number of fusions: $4(n\cdot k)\ L^3$
    \item Number of optical elements needed per fusion (success rate of 75\%): 15 polarization rotators and 4 polarising beamsplitters
\end{itemize}

The variables that correlate with the size of the computer/computation are $n$ and $k$ as we don't expect to change the encoding ($L$). It might be the case that for some small computations we choose a smaller renormalised qubit to save resources, but for big computations, choosing a renormalised qubit of $L=6,7,8$ should be good enough for all, and can therefore be taken as a constant. Therefore as the size of the computation only affects variable $n$ and $k$, we can take it as if $(n\cdot k)$ was the size of the computation and the dependance of the resources on this variable is linear: given on demand 3-photon GHZ and Bell pairs, the scaling is \emph{linear} on the size of the computer.

\section{Comparison with previous percolation schemes}

When designing a feasible architecture for quantum computing, the size of the machine (in terms of number of components and resources required) is one of the biggest concerns. The following comparison shows that our design utilises at least an order of magnitude fewer resources than the design of Kieling et al. To show this we extract data presented in their paper and compared it with ours in the same conditions.

In figure 4 of their paper \cite{Kieling2007}, the authors show the dependance of the diamond lattice block size $k^3$ on the size $L$ of the renormalized square lattice for three different sets of site bond probabilities $(p_{site},p_{bond})$. The overall success probability threshold $P(L)$ was chosen to be $\frac{1}{2}$. In our work, we have performed all the simulations, assuming the GHZ states are provided deterministically. For the scheme comparison to be fair, we choose to compare with the data points that correspond with the data set $(1.00,0.5)$. From the data used to produce figure 7 in our paper, we extract for different $k$s, what is the maximum value of $L$ we can reach with $\Pi(L)\ge\frac{1}{2}$.

\begin{figure}[hbt]
    \begin{center}
  \includegraphics[width=0.5\linewidth]{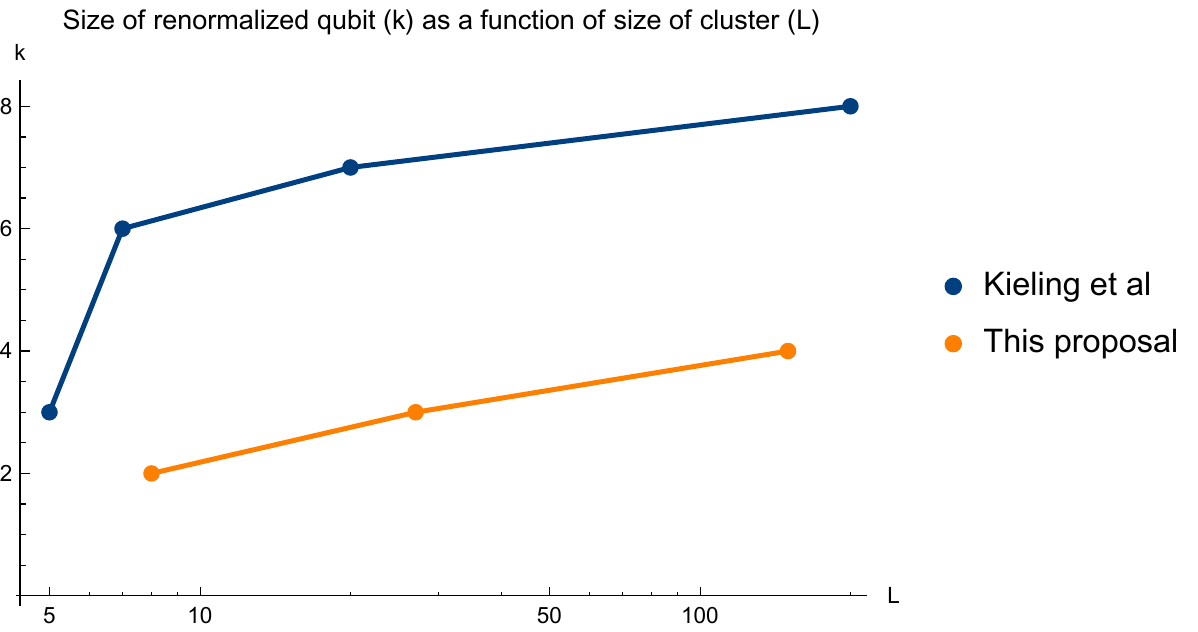}
  \caption{Comparison of the size of the renormalised qubit $(k)$ for different cluster sizes $(L)$}
  \label{fig:comparison1}
    \end{center}
\end{figure}

We can see in figure \ref{fig:comparison1} that already there is a significant improvement in our scheme, as the size of the renormalised qubit is reduces noticeably in our scheme with respect to Kieling etc al scheme \cite{Kieling2007}. But the improvement becomes much greater once we consider the number of Bell pairs that are needed to build each renormalised qubit and the entire cluster.

To obtain this comparison, we will first calculate how many Bell pairs are needed to obtain a GHZ with probability greater than 99.9999\%.

\begin{itemize}
	\item The data in \cite{Kieling2007} is obtained for 4-photon GHZ states. For each 4-photon GHZ state we need 3 Bell pairs and the LON works with probability $\frac{1}{4}$. In order to have a deterministic 4-photon GHZ  we must repeat the generation procedure $t$ times, where $t$ is $1-(1-\frac{1}{4}))^t \ge 1 \Rightarrow t=51$. In total we consumer $3\times 51=153$ Bell pairs in the generation of a deterministic 4-photon GHZ state.
	\item In our paper we require deterministic 3-photon GHZ states. For each attempt at generating one, we need 2 Bell pairs and  the LON works with probability of success $\frac{1}{2}$. In order to have a deterministic 3-photon GHZ we must repeat the generation procedure $t$ times, where $t$ is $1-(1-\frac{1}{2}))^t \ge 1 \Rightarrow t=21$. In total we consumer $2\times 21=42$ Bell pairs in the generation of a deterministic 3-photon GHZ state.
\end{itemize}

\begin{figure}[hbt]
    \begin{center}
  \includegraphics[width=0.5\linewidth]{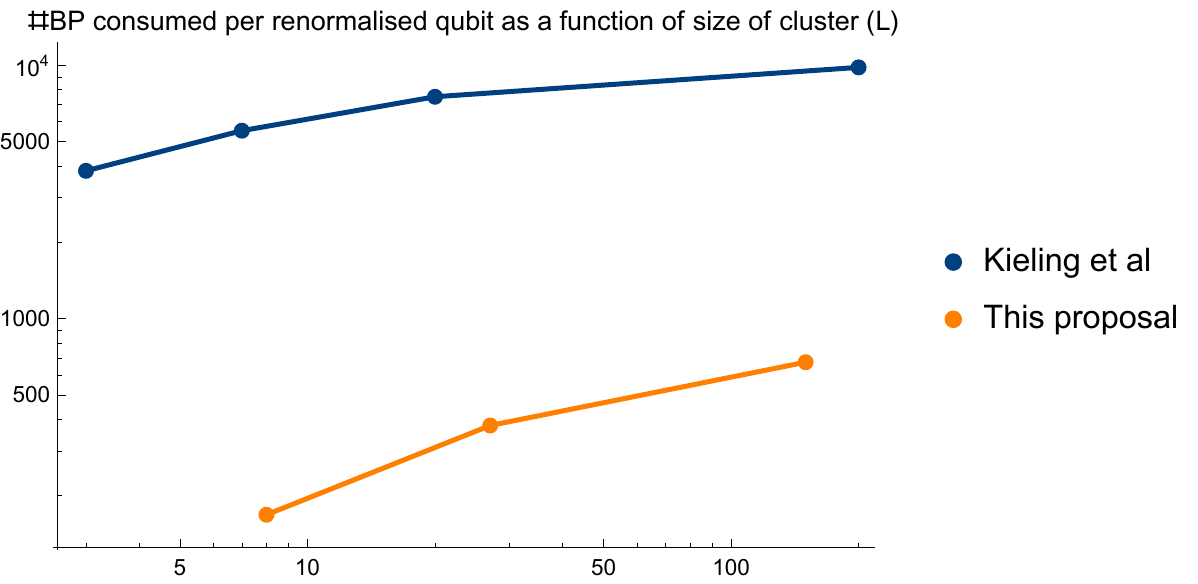}
  \caption{Comparison of the number of Bell pairs consumed per renormalised qubits for different cluster sizes $(L)$}
  \label{fig:comparison2}
    \end{center}
\end{figure}

\begin{figure}[hbt]
    \begin{center}
  \includegraphics[width=0.5\linewidth]{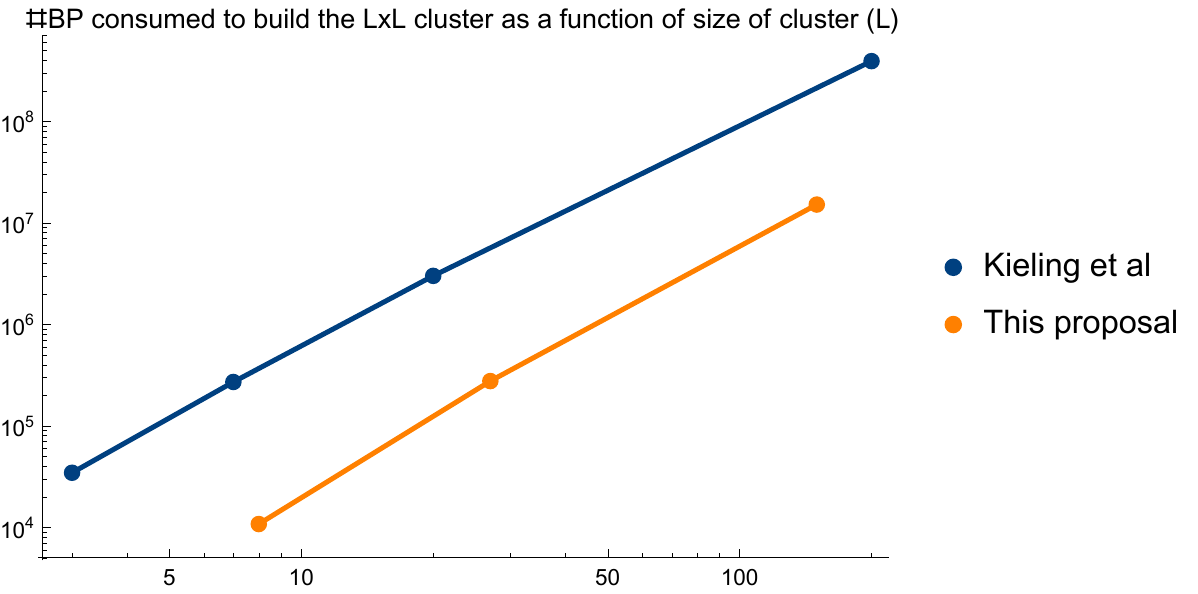}
  \caption{Comparison of the number of Bell pairs consumed to build the entire $L\times L$ cluster for different cluster sizes $(L)$}
  \label{fig:comparison3}
    \end{center}
\end{figure}

With these numbers, we can transform the data in figure \ref{fig:comparison1} into number of consumable resources used for different cluster sizes. In figure \ref{fig:comparison2} we can see the number of Bell pairs consumed per renormalized qubit and in figure \ref{fig:comparison3}, the number of Bell pairs required to build an $L \times L$ lattice of renormalized qubits. We can clearly see from figure \ref{fig:comparison3} that the resources required to build a renormalised cluster state are an order of magnitude smaller in our proposal in comparison with the scheme presented in \cite{Kieling2007}.


To provide a quantitative resource comparison of the data presented in figures 4, 5, 6, we calculate the ratio between the number of Bell pairs needed to build a cluster state of a similar size for both schemes. Comparing points with L of the same order of magnitude, we find that \cite{Kieling2007} uses at least 14\% more Bell pairs than our scheme to build the cluster state. We want to emphasise that our scheme offers further benefits in addition to this reduction in resources. The main differences between this proposal and \cite{Kieling2007} are the size of the resource states used to build the cluster state and the type of gate that is applied to these resource states in order to built the cluster state. In terms of resources, figures \ref{fig:comparison2} and \ref{fig:comparison3} clearly show that building the cluster out of smaller GHZ states consumes less resources overall, the number of Bell pairs needed to probabilistically create the GHZ is lower and the probability of success is higher. The type of gate used in \cite{Kieling2007},  is the Type-I gate introduced in \cite{Browne2005}, this gate only measures one photon out of the two photons that go into the gate. A failure of the gate combined with the loss of a photon would produce a false positive, inducing a logical error on the cluster \cite{Varnava2008} . In our proposal we only use the Type-II boosted gate which we introduce in this manuscript. This is a loss tolerant gate as it measures all the photons involved, and therefore it can never be the case that a loss of a photon transforms into a logical error. Therefore, the improvements of our scheme over the scheme presented by Kieling et al. are not only the reduction of the amount of resources needed, but also on the overall robustness of the construction, which is indicated by the 5\% improvement on the heralded loss tolerance.


\begin{thebibliography}{27}%
\makeatletter
\providecommand \@ifxundefined [1]{%
 \@ifx{#1\undefined}
}%
\providecommand \@ifnum [1]{%
 \ifnum #1\expandafter \@firstoftwo
 \else \expandafter \@secondoftwo
 \fi
}%
\providecommand \@ifx [1]{%
 \ifx #1\expandafter \@firstoftwo
 \else \expandafter \@secondoftwo
 \fi
}%
\providecommand \natexlab [1]{#1}%
\providecommand \enquote  [1]{``#1''}%
\providecommand \bibnamefont  [1]{#1}%
\providecommand \bibfnamefont [1]{#1}%
\providecommand \citenamefont [1]{#1}%
\providecommand \href@noop [0]{\@secondoftwo}%
\providecommand \href [0]{\begingroup \@sanitize@url \@href}%
\providecommand \@href[1]{\@@startlink{#1}\@@href}%
\providecommand \@@href[1]{\endgroup#1\@@endlink}%
\providecommand \@sanitize@url [0]{\catcode `\\12\catcode `\$12\catcode
  `\&12\catcode `\#12\catcode `\^12\catcode `\_12\catcode `\%12\relax}%
\providecommand \@@startlink[1]{}%
\providecommand \@@endlink[0]{}%
\providecommand \url  [0]{\begingroup\@sanitize@url \@url }%
\providecommand \@url [1]{\endgroup\@href {#1}{\urlprefix }}%
\providecommand \urlprefix  [0]{URL }%
\providecommand \Eprint [0]{\href }%
\@ifxundefined \urlstyle {%
  \providecommand \doi  [0]{\begingroup \@sanitize@url \@doi}%
  \providecommand \@doi [1]{\endgroup \@@startlink {\doibase
  #1}doi:\discretionary {}{}{}#1\@@endlink }%
}{%
  \providecommand \doi  [0]{doi:\discretionary{}{}{}\begingroup
  \urlstyle{rm}\Url }%
}%
\providecommand \doibase [0]{http://dx.doi.org/}%
\providecommand \Doi [0]{\begingroup \@sanitize@url \@Doi }%
\providecommand \@Doi  [1]{\endgroup\@@startlink{\doibase#1}\@@Doi}%
\providecommand \@@Doi [1]{#1\@@endlink}%
\providecommand \selectlanguage [0]{\@gobble}%
\providecommand \bibinfo  [0]{\@secondoftwo}%
\providecommand \bibfield  [0]{\@secondoftwo}%
\providecommand \translation [1]{[#1]}%
\providecommand \BibitemOpen [0]{}%
\providecommand \bibitemStop [0]{}%
\providecommand \bibitemNoStop [0]{.\EOS\space}%
\providecommand \EOS [0]{\spacefactor3000\relax}%
\providecommand \BibitemShut  [1]{\csname bibitem#1\endcsname}%
\bibitem [{\citenamefont {Knill}\ \emph {et~al.}(2001)\citenamefont {Knill},
  \citenamefont {Laflamme},\ and\ \citenamefont {Milburn}}]{Knill1998}%
  \BibitemOpen
  \bibfield  {author} {\bibinfo {author} {\bibfnamefont {E.}~\bibnamefont
  {Knill}}, \bibinfo {author} {\bibfnamefont {R.}~\bibnamefont {Laflamme}}, \
  and\ \bibinfo {author} {\bibfnamefont {G.~J.}\ \bibnamefont {Milburn}},\
  }\href@noop {} {\bibfield  {journal} {\bibinfo  {journal} {Nature},\ }\textbf
  {\bibinfo {volume} {409}},\ \bibinfo {pages} {46} (\bibinfo {year}
  {2001})}\BibitemShut {NoStop}%
\bibitem [{\citenamefont {Nielsen}(2004)}]{Nielsen2004}%
  \BibitemOpen
  \bibfield  {author} {\bibinfo {author} {\bibfnamefont {M.~A.}\ \bibnamefont
  {Nielsen}},\ }\Doi {10.1103/PhysRevLett.93.040503} {\bibfield  {journal}
  {\bibinfo  {journal} {Physical Review Letters},\ }\textbf {\bibinfo {volume}
  {93}},\ \bibinfo {pages} {040503} (\bibinfo {year} {2004})}\BibitemShut
  {NoStop}%
\bibitem [{\citenamefont {Browne}\ and\ \citenamefont
  {Rudolph}(2005)}]{Browne2005}%
  \BibitemOpen
  \bibfield  {author} {\bibinfo {author} {\bibfnamefont {D.~E.}\ \bibnamefont
  {Browne}}\ and\ \bibinfo {author} {\bibfnamefont {T.}~\bibnamefont
  {Rudolph}},\ }\Doi {10.1103/PhysRevLett.95.010501} {\bibfield  {journal}
  {\bibinfo  {journal} {Phys. Rev. Lett.},\ }\textbf {\bibinfo {volume} {95}},\
  \bibinfo {pages} {010501} (\bibinfo {year} {2005})}\BibitemShut {NoStop}%
\bibitem [{\citenamefont {Kieling}\ \emph {et~al.}(2007)\citenamefont
  {Kieling}, \citenamefont {Rudolph},\ and\ \citenamefont
  {Eisert}}]{Kieling2007}%
  \BibitemOpen
  \bibfield  {author} {\bibinfo {author} {\bibfnamefont {K.}~\bibnamefont
  {Kieling}}, \bibinfo {author} {\bibfnamefont {T.}~\bibnamefont {Rudolph}}, \
  and\ \bibinfo {author} {\bibfnamefont {J.}~\bibnamefont {Eisert}},\ }\Doi
  {10.1103/PhysRevLett.99.130501} {\bibfield  {journal} {\bibinfo  {journal}
  {Physical Review Letters},\ }\textbf {\bibinfo {volume} {99}},\ \bibinfo
  {pages} {130501} (\bibinfo {year} {2007})}\BibitemShut {NoStop}%
\bibitem [{\citenamefont {Kok}\ \emph {et~al.}(2007)\citenamefont {Kok},
  \citenamefont {Nemoto}, \citenamefont {Ralph}, \citenamefont {Dowling},\ and\
  \citenamefont {Milburn}}]{Kok2007}%
  \BibitemOpen
  \bibfield  {author} {\bibinfo {author} {\bibfnamefont {P.}~\bibnamefont
  {Kok}}, \bibinfo {author} {\bibfnamefont {K.}~\bibnamefont {Nemoto}},
  \bibinfo {author} {\bibfnamefont {T.~C.}\ \bibnamefont {Ralph}}, \bibinfo
  {author} {\bibfnamefont {J.~P.}\ \bibnamefont {Dowling}}, \ and\ \bibinfo
  {author} {\bibfnamefont {G.~J.}\ \bibnamefont {Milburn}},\ }\Doi
  {10.1103/RevModPhys.79.135} {\bibfield  {journal} {\bibinfo  {journal}
  {Reviews of Modern Physics},\ }\textbf {\bibinfo {volume} {79}},\ \bibinfo
  {pages} {135} (\bibinfo {year} {2007})}\BibitemShut {NoStop}%
\bibitem [{\citenamefont {Hayes}\ \emph {et~al.}(2010)\citenamefont {Hayes},
  \citenamefont {Haselgrove}, \citenamefont {Gilchrist},\ and\ \citenamefont
  {Ralph}}]{Hayes2010}%
  \BibitemOpen
  \bibfield  {author} {\bibinfo {author} {\bibfnamefont {A.~J.~F.}\
  \bibnamefont {Hayes}}, \bibinfo {author} {\bibfnamefont {H.~L.}\ \bibnamefont
  {Haselgrove}}, \bibinfo {author} {\bibfnamefont {A.}~\bibnamefont
  {Gilchrist}}, \ and\ \bibinfo {author} {\bibfnamefont {T.~C.}\ \bibnamefont
  {Ralph}},\ }\Doi {10.1103/PhysRevA.82.022323} {\bibfield  {journal} {\bibinfo
   {journal} {Physical Review A},\ }\textbf {\bibinfo {volume} {82}},\ \bibinfo
  {pages} {022323} (\bibinfo {year} {2010})}\BibitemShut {NoStop}%
\bibitem [{\citenamefont {Lund}\ \emph {et~al.}(2008)\citenamefont {Lund},
  \citenamefont {Ralph},\ and\ \citenamefont {Haselgrove}}]{Lund2008}%
  \BibitemOpen
  \bibfield  {author} {\bibinfo {author} {\bibfnamefont {A.~P.}\ \bibnamefont
  {Lund}}, \bibinfo {author} {\bibfnamefont {T.~C.}\ \bibnamefont {Ralph}}, \
  and\ \bibinfo {author} {\bibfnamefont {H.~L.}\ \bibnamefont {Haselgrove}},\
  }\Doi {10.1103/PhysRevLett.100.030503} {\bibfield  {journal} {\bibinfo
  {journal} {Physical Review Letters},\ }\textbf {\bibinfo {volume} {100}},\
  \bibinfo {pages} {030503} (\bibinfo {year} {2008})}\BibitemShut {NoStop}%
\bibitem [{\citenamefont {{Carolan}}\ \emph {et~al.}(2014)\citenamefont
  {{Carolan}}, \citenamefont {{Meinecke}}, \citenamefont {{Shadbolt}},
  \citenamefont {{Russell}}, \citenamefont {{Ismail}}, \citenamefont
  {{W{\"o}rhoff}}, \citenamefont {{Rudolph}}, \citenamefont {{Thompson}},
  \citenamefont {{O'Brien}}, \citenamefont {{Matthews}},\ and\ \citenamefont
  {{Laing}}}]{carolan2014}%
  \BibitemOpen
  \bibfield  {author} {\bibinfo {author} {\bibfnamefont {J.}~\bibnamefont
  {{Carolan}}}, \bibinfo {author} {\bibfnamefont {J.~D.~A.}\ \bibnamefont
  {{Meinecke}}}, \bibinfo {author} {\bibfnamefont {P.~J.}\ \bibnamefont
  {{Shadbolt}}}, \bibinfo {author} {\bibfnamefont {N.~J.}\ \bibnamefont
  {{Russell}}}, \bibinfo {author} {\bibfnamefont {N.}~\bibnamefont {{Ismail}}},
  \bibinfo {author} {\bibfnamefont {K.}~\bibnamefont {{W{\"o}rhoff}}}, \bibinfo
  {author} {\bibfnamefont {T.}~\bibnamefont {{Rudolph}}}, \bibinfo {author}
  {\bibfnamefont {M.~G.}\ \bibnamefont {{Thompson}}}, \bibinfo {author}
  {\bibfnamefont {J.~L.}\ \bibnamefont {{O'Brien}}}, \bibinfo {author}
  {\bibfnamefont {J.~C.~F.}\ \bibnamefont {{Matthews}}}, \ and\ \bibinfo
  {author} {\bibfnamefont {A.}~\bibnamefont {{Laing}}},\ }\Doi
  {10.1038/nphoton.2014.152} {\bibfield  {journal} {\bibinfo  {journal} {Nature
  Photonics},\ }\textbf {\bibinfo {volume} {8}},\ \bibinfo {pages} {621}
  (\bibinfo {year} {2014})},\ \Eprint {http://arxiv.org/abs/1311.2913}
  {arXiv:1311.2913 [quant-ph]} \BibitemShut {NoStop}%
\bibitem [{\citenamefont {{Silverstone}}\ \emph {et~al.}(2014)\citenamefont
  {{Silverstone}}, \citenamefont {{Santagati}}, \citenamefont {{Bonneau}},
  \citenamefont {{Strain}}, \citenamefont {{Sorel}}, \citenamefont
  {{O'Brien}},\ and\ \citenamefont {{Thompson}}}]{silverstone2014}%
  \BibitemOpen
  \bibfield  {author} {\bibinfo {author} {\bibfnamefont {J.~W.}\ \bibnamefont
  {{Silverstone}}}, \bibinfo {author} {\bibfnamefont {R.}~\bibnamefont
  {{Santagati}}}, \bibinfo {author} {\bibfnamefont {D.}~\bibnamefont
  {{Bonneau}}}, \bibinfo {author} {\bibfnamefont {M.~J.}\ \bibnamefont
  {{Strain}}}, \bibinfo {author} {\bibfnamefont {M.}~\bibnamefont {{Sorel}}},
  \bibinfo {author} {\bibfnamefont {J.~L.}\ \bibnamefont {{O'Brien}}}, \ and\
  \bibinfo {author} {\bibfnamefont {M.~G.}\ \bibnamefont {{Thompson}}},\
  }\href@noop {} {\bibfield  {journal} {\bibinfo  {journal} {ArXiv e-prints}}
  (\bibinfo {year} {2014})},\ \Eprint {http://arxiv.org/abs/1410.8332}
  {arXiv:1410.8332 [quant-ph]} \BibitemShut {NoStop}%
\bibitem [{\citenamefont {{Spagnolo}}\ \emph {et~al.}(2014)\citenamefont
  {{Spagnolo}}, \citenamefont {{Vitelli}}, \citenamefont {{Bentivegna}},
  \citenamefont {{Brod}}, \citenamefont {{Crespi}}, \citenamefont {{Flamini}},
  \citenamefont {{Giacomini}}, \citenamefont {{Milani}}, \citenamefont
  {{Ramponi}}, \citenamefont {{Mataloni}}, \citenamefont {{Osellame}},
  \citenamefont {{Galv{\~a}o}},\ and\ \citenamefont
  {{Sciarrino}}}]{spagnolo2014}%
  \BibitemOpen
  \bibfield  {author} {\bibinfo {author} {\bibfnamefont {N.}~\bibnamefont
  {{Spagnolo}}}, \bibinfo {author} {\bibfnamefont {C.}~\bibnamefont
  {{Vitelli}}}, \bibinfo {author} {\bibfnamefont {M.}~\bibnamefont
  {{Bentivegna}}}, \bibinfo {author} {\bibfnamefont {D.~J.}\ \bibnamefont
  {{Brod}}}, \bibinfo {author} {\bibfnamefont {A.}~\bibnamefont {{Crespi}}},
  \bibinfo {author} {\bibfnamefont {F.}~\bibnamefont {{Flamini}}}, \bibinfo
  {author} {\bibfnamefont {S.}~\bibnamefont {{Giacomini}}}, \bibinfo {author}
  {\bibfnamefont {G.}~\bibnamefont {{Milani}}}, \bibinfo {author}
  {\bibfnamefont {R.}~\bibnamefont {{Ramponi}}}, \bibinfo {author}
  {\bibfnamefont {P.}~\bibnamefont {{Mataloni}}}, \bibinfo {author}
  {\bibfnamefont {R.}~\bibnamefont {{Osellame}}}, \bibinfo {author}
  {\bibfnamefont {E.~F.}\ \bibnamefont {{Galv{\~a}o}}}, \ and\ \bibinfo
  {author} {\bibfnamefont {F.}~\bibnamefont {{Sciarrino}}},\ }\Doi
  {10.1038/nphoton.2014.135} {\bibfield  {journal} {\bibinfo  {journal} {Nature
  Photonics},\ }\textbf {\bibinfo {volume} {8}},\ \bibinfo {pages} {615}
  (\bibinfo {year} {2014})},\ \Eprint {http://arxiv.org/abs/1311.1622}
  {arXiv:1311.1622 [quant-ph]} \BibitemShut {NoStop}%
\bibitem [{\citenamefont {{Cai}}\ \emph {et~al.}(2014)\citenamefont {{Cai}},
  \citenamefont {{Wu}}, \citenamefont {{Su}}, \citenamefont {{Chen}},
  \citenamefont {{Wang}}, \citenamefont {{Li}}, \citenamefont {{Liu}},
  \citenamefont {{Lu}},\ and\ \citenamefont {{Pan}}}]{cai2014}%
  \BibitemOpen
  \bibfield  {author} {\bibinfo {author} {\bibfnamefont {X.-D.}\ \bibnamefont
  {{Cai}}}, \bibinfo {author} {\bibfnamefont {D.}~\bibnamefont {{Wu}}},
  \bibinfo {author} {\bibfnamefont {Z.-E.}\ \bibnamefont {{Su}}}, \bibinfo
  {author} {\bibfnamefont {M.-C.}\ \bibnamefont {{Chen}}}, \bibinfo {author}
  {\bibfnamefont {X.-L.}\ \bibnamefont {{Wang}}}, \bibinfo {author}
  {\bibfnamefont {L.}~\bibnamefont {{Li}}}, \bibinfo {author} {\bibfnamefont
  {N.-L.}\ \bibnamefont {{Liu}}}, \bibinfo {author} {\bibfnamefont {C.-Y.}\
  \bibnamefont {{Lu}}}, \ and\ \bibinfo {author} {\bibfnamefont {J.-W.}\
  \bibnamefont {{Pan}}},\ }\href@noop {} {\bibfield  {journal} {\bibinfo
  {journal} {ArXiv e-prints}} (\bibinfo {year} {2014})},\ \Eprint
  {http://arxiv.org/abs/1409.7770} {arXiv:1409.7770 [quant-ph]} \BibitemShut
  {NoStop}%
\bibitem [{\citenamefont {{Barz}}\ \emph {et~al.}(2014)\citenamefont {{Barz}},
  \citenamefont {{Vasconcelos}}, \citenamefont {{Greganti}}, \citenamefont
  {{Zwerger}}, \citenamefont {{D{\"u}r}}, \citenamefont {{Briegel}},\ and\
  \citenamefont {{Walther}}}]{barz2014}%
  \BibitemOpen
  \bibfield  {author} {\bibinfo {author} {\bibfnamefont {S.}~\bibnamefont
  {{Barz}}}, \bibinfo {author} {\bibfnamefont {R.}~\bibnamefont
  {{Vasconcelos}}}, \bibinfo {author} {\bibfnamefont {C.}~\bibnamefont
  {{Greganti}}}, \bibinfo {author} {\bibfnamefont {M.}~\bibnamefont
  {{Zwerger}}}, \bibinfo {author} {\bibfnamefont {W.}~\bibnamefont
  {{D{\"u}r}}}, \bibinfo {author} {\bibfnamefont {H.~J.}\ \bibnamefont
  {{Briegel}}}, \ and\ \bibinfo {author} {\bibfnamefont {P.}~\bibnamefont
  {{Walther}}},\ }\Doi {10.1103/PhysRevA.90.042302} {\bibfield  {journal}
  {\bibinfo  {journal} {\pra},\ }\textbf {\bibinfo {volume} {90}},\ \bibinfo
  {eid} {042302} (\bibinfo {year} {2014})},\ \Eprint
  {http://arxiv.org/abs/1308.5209} {arXiv:1308.5209 [quant-ph]} \BibitemShut
  {NoStop}%
\bibitem [{\citenamefont {Grice}(2011)}]{Grice2011}%
  \BibitemOpen
  \bibfield  {author} {\bibinfo {author} {\bibfnamefont {W.~P.}\ \bibnamefont
  {Grice}},\ }\Doi {10.1103/PhysRevA.84.042331} {\bibfield  {journal} {\bibinfo
   {journal} {Physical Review A},\ }\textbf {\bibinfo {volume} {84}},\ \bibinfo
  {pages} {042331} (\bibinfo {year} {2011})}\BibitemShut {NoStop}%
\bibitem [{\citenamefont {Ewert}\ and\ \citenamefont {van
  Loock}(2014)}]{Ewert2014}%
  \BibitemOpen
  \bibfield  {author} {\bibinfo {author} {\bibfnamefont {F.}~\bibnamefont
  {Ewert}}\ and\ \bibinfo {author} {\bibfnamefont {P.}~\bibnamefont {van
  Loock}},\ }\Doi {10.1103/PhysRevLett.113.140403} {\bibfield  {journal}
  {\bibinfo  {journal} {Phys. Rev. Lett.},\ }\textbf {\bibinfo {volume}
  {113}},\ \bibinfo {pages} {140403} (\bibinfo {year} {2014})}\BibitemShut
  {NoStop}%
\bibitem [{Note1()}]{Note1}%
  \BibitemOpen
  \bibinfo {note} {The best known theoretical GHZ building strategy from Bell
  Pairs has a success probability $p_{succ}=\left (\protect \frac {1}{2}\right
  )^{\delimiter "4262304 \protect \frac {n-1}{2}\delimiter "5263305 }\left
  (\protect \frac {3}{4}\right )^{\delimiter "4264306 \protect \frac
  {n-1}{2}\delimiter "5265307 }$. Gimeno-Segovia et al. (in
  preparation)}\BibitemShut {NoStop}%
\bibitem [{\citenamefont {Calsamiglia}\ and\ \citenamefont
  {L{\"u}tkenhaus}(2001)}]{calsamiglia2001maximum}%
  \BibitemOpen
  \bibfield  {author} {\bibinfo {author} {\bibfnamefont {J.}~\bibnamefont
  {Calsamiglia}}\ and\ \bibinfo {author} {\bibfnamefont {N.}~\bibnamefont
  {L{\"u}tkenhaus}},\ }\href@noop {} {\bibfield  {journal} {\bibinfo  {journal}
  {Applied Physics B},\ }\textbf {\bibinfo {volume} {72}},\ \bibinfo {pages}
  {67} (\bibinfo {year} {2001})}\BibitemShut {NoStop}%
\bibitem [{\citenamefont {Varnava}\ \emph {et~al.}(2008)\citenamefont
  {Varnava}, \citenamefont {Browne},\ and\ \citenamefont
  {Rudolph}}]{Varnava2008}%
  \BibitemOpen
  \bibfield  {author} {\bibinfo {author} {\bibfnamefont {M.}~\bibnamefont
  {Varnava}}, \bibinfo {author} {\bibfnamefont {D.~E.}\ \bibnamefont {Browne}},
  \ and\ \bibinfo {author} {\bibfnamefont {T.}~\bibnamefont {Rudolph}},\ }\Doi
  {10.1103/PhysRevLett.100.060502} {\bibfield  {journal} {\bibinfo  {journal}
  {Physical Review Letters},\ }\textbf {\bibinfo {volume} {100}},\ \bibinfo
  {pages} {060502} (\bibinfo {year} {2008})}\BibitemShut {NoStop}%
\bibitem [{\citenamefont {Stauffer}\ and\ \citenamefont
  {Aharony}(1994)}]{Stauffer1994}%
  \BibitemOpen
  \bibfield  {author} {\bibinfo {author} {\bibfnamefont {D.}~\bibnamefont
  {Stauffer}}\ and\ \bibinfo {author} {\bibfnamefont {A.}~\bibnamefont
  {Aharony}},\ }\Doi {10.1103/RevModPhys.63.991} {\emph {\bibinfo {title}
  {Introduction to Percolation Theory}}},\ Vol.~\bibinfo {volume} {1}\
  (\bibinfo {year} {1994})\ ISBN \bibinfo {isbn} {0748402535},\ p.\ \bibinfo
  {pages} {192}\BibitemShut {NoStop}%
\bibitem [{\citenamefont {Kieling}\ and\ \citenamefont
  {Eisert}()}]{Kieling2007a}%
  \BibitemOpen
  \bibfield  {author} {\bibinfo {author} {\bibfnamefont {K.}~\bibnamefont
  {Kieling}}\ and\ \bibinfo {author} {\bibfnamefont {J.}~\bibnamefont
  {Eisert}},\ }in\ \href {http://arxiv.org/abs/0712.1836} {\emph {\bibinfo
  {booktitle} {Quantum and Semi-classical Percolation and Breakdown in
  Disordered Solids, (Springer, Berlin, 2009)}}},\ pp.\ \bibinfo {pages}
  {287--319}\BibitemShut {NoStop}%
\bibitem [{\citenamefont {Tarasevich}\ and\ \citenamefont {van~der
  Marck}(1999)}]{Tarasevich1999}%
  \BibitemOpen
  \bibfield  {author} {\bibinfo {author} {\bibfnamefont {Y.~Y.}\ \bibnamefont
  {Tarasevich}}\ and\ \bibinfo {author} {\bibfnamefont {S.~C.}\ \bibnamefont
  {van~der Marck}},\ }\href
  {http://www.worldscientific.com/doi/pdf/10.1142/S0129183199000978
  http://arxiv.org/abs/cond-mat/9906078} {\bibfield  {journal} {\bibinfo
  {journal} {International Journal of Modern Physics C},\ }\textbf {\bibinfo
  {volume} {10}},\ \bibinfo {pages} {14} (\bibinfo {year} {1999})}\BibitemShut
  {NoStop}%
\bibitem [{\citenamefont {Varnava}\ \emph {et~al.}(2006)\citenamefont
  {Varnava}, \citenamefont {Browne},\ and\ \citenamefont
  {Rudolph}}]{treecluster}%
  \BibitemOpen
  \bibfield  {author} {\bibinfo {author} {\bibfnamefont {M.}~\bibnamefont
  {Varnava}}, \bibinfo {author} {\bibfnamefont {D.~E.}\ \bibnamefont {Browne}},
  \ and\ \bibinfo {author} {\bibfnamefont {T.}~\bibnamefont {Rudolph}},\ }\Doi
  {10.1103/PhysRevLett.97.120501} {\bibfield  {journal} {\bibinfo  {journal}
  {Phys. Rev. Lett.},\ }\textbf {\bibinfo {volume} {97}},\ \bibinfo {pages}
  {120501} (\bibinfo {year} {2006})}\BibitemShut {NoStop}%
\bibitem [{\citenamefont {Bravyi}\ and\ \citenamefont
  {Kitaev}(1998)}]{surface}%
  \BibitemOpen
  \bibfield  {author} {\bibinfo {author} {\bibfnamefont {S.~B.}\ \bibnamefont
  {Bravyi}}\ and\ \bibinfo {author} {\bibfnamefont {A.~Y.}\ \bibnamefont
  {Kitaev}},\ }\href@noop {} {\bibfield  {journal} {\bibinfo  {journal} {arXiv
  preprint quant-ph/9811052}} (\bibinfo {year} {1998})}\BibitemShut {NoStop}%
\bibitem [{\citenamefont {Dennis}\ \emph {et~al.}(2002)\citenamefont {Dennis},
  \citenamefont {Kitaev}, \citenamefont {Landahl},\ and\ \citenamefont
  {Preskill}}]{topological}%
  \BibitemOpen
  \bibfield  {author} {\bibinfo {author} {\bibfnamefont {E.}~\bibnamefont
  {Dennis}}, \bibinfo {author} {\bibfnamefont {A.}~\bibnamefont {Kitaev}},
  \bibinfo {author} {\bibfnamefont {A.}~\bibnamefont {Landahl}}, \ and\
  \bibinfo {author} {\bibfnamefont {J.}~\bibnamefont {Preskill}},\ }\href@noop
  {} {\bibfield  {journal} {\bibinfo  {journal} {Journal of Mathematical
  Physics},\ }\textbf {\bibinfo {volume} {43}},\ \bibinfo {pages} {4452}
  (\bibinfo {year} {2002})}\BibitemShut {NoStop}%
\bibitem [{\citenamefont {Raussendorf}\ \emph {et~al.}(2006)\citenamefont
  {Raussendorf}, \citenamefont {Harrington},\ and\ \citenamefont
  {Goyal}}]{raussendorf2006}%
  \BibitemOpen
  \bibfield  {author} {\bibinfo {author} {\bibfnamefont {R.}~\bibnamefont
  {Raussendorf}}, \bibinfo {author} {\bibfnamefont {J.}~\bibnamefont
  {Harrington}}, \ and\ \bibinfo {author} {\bibfnamefont {K.}~\bibnamefont
  {Goyal}},\ }\href@noop {} {\bibfield  {journal} {\bibinfo  {journal} {Annals
  of physics},\ }\textbf {\bibinfo {volume} {321}},\ \bibinfo {pages} {2242}
  (\bibinfo {year} {2006})}\BibitemShut {NoStop}%
\bibitem [{\citenamefont {Bonneau}\ \emph {et~al.}(2014)\citenamefont
  {Bonneau}, \citenamefont {Mendoza}, \citenamefont {O'Brien},\ and\
  \citenamefont {Thompson}}]{Bonneau2014}%
  \BibitemOpen
  \bibfield  {author} {\bibinfo {author} {\bibfnamefont {D.}~\bibnamefont
  {Bonneau}}, \bibinfo {author} {\bibfnamefont {G.~J.}\ \bibnamefont
  {Mendoza}}, \bibinfo {author} {\bibfnamefont {J.~L.}\ \bibnamefont
  {O'Brien}}, \ and\ \bibinfo {author} {\bibfnamefont {M.~G.}\ \bibnamefont
  {Thompson}},\ }\href {http://arxiv.org/abs/1409.5341} {\bibfield  {journal}
  {\bibinfo  {journal} {arXiv preprint 1409.5341}} (\bibinfo {year}
  {2014})}\BibitemShut {NoStop}%
\bibitem [{\citenamefont {Lindner}\ and\ \citenamefont
  {Rudolph}(2009)}]{Lindner2009}%
  \BibitemOpen
  \bibfield  {author} {\bibinfo {author} {\bibfnamefont {N.~H.}\ \bibnamefont
  {Lindner}}\ and\ \bibinfo {author} {\bibfnamefont {T.}~\bibnamefont
  {Rudolph}},\ }\Doi {10.1103/PhysRevLett.103.113602} {\bibfield  {journal}
  {\bibinfo  {journal} {Physical Review Letters},\ }\textbf {\bibinfo {volume}
  {103}},\ \bibinfo {pages} {113602} (\bibinfo {year} {2009})}\BibitemShut
  {NoStop}%
\bibitem [{\citenamefont {Zaidi}\ \emph {et~al.}(2015)\citenamefont {Zaidi},
  \citenamefont {Dawson}, \citenamefont {van Loock},\ and\ \citenamefont
  {Rudolph}}]{Zaidi2014}%
  \BibitemOpen
  \bibfield  {author} {\bibinfo {author} {\bibfnamefont {H.~A.}\ \bibnamefont
  {Zaidi}}, \bibinfo {author} {\bibfnamefont {C.}~\bibnamefont {Dawson}},
  \bibinfo {author} {\bibfnamefont {P.}~\bibnamefont {van Loock}}, \ and\
  \bibinfo {author} {\bibfnamefont {T.}~\bibnamefont {Rudolph}},\ }\Doi
  {10.1103/PhysRevA.91.042301} {\bibfield  {journal} {\bibinfo  {journal}
  {Phys. Rev. A},\ }\textbf {\bibinfo {volume} {91}},\ \bibinfo {pages}
  {042301} (\bibinfo {year} {2015})}\BibitemShut {NoStop}%
\end{thebibliography}

\begin{thebibliography}{6}%
\makeatletter
\providecommand \@ifxundefined [1]{%
 \@ifx{#1\undefined}
}%
\providecommand \@ifnum [1]{%
 \ifnum #1\expandafter \@firstoftwo
 \else \expandafter \@secondoftwo
 \fi
}%
\providecommand \@ifx [1]{%
 \ifx #1\expandafter \@firstoftwo
 \else \expandafter \@secondoftwo
 \fi
}%
\providecommand \natexlab [1]{#1}%
\providecommand \enquote  [1]{``#1''}%
\providecommand \bibnamefont  [1]{#1}%
\providecommand \bibfnamefont [1]{#1}%
\providecommand \citenamefont [1]{#1}%
\providecommand \href@noop [0]{\@secondoftwo}%
\providecommand \href [0]{\begingroup \@sanitize@url \@href}%
\providecommand \@href[1]{\@@startlink{#1}\@@href}%
\providecommand \@@href[1]{\endgroup#1\@@endlink}%
\providecommand \@sanitize@url [0]{\catcode `\\12\catcode `\$12\catcode
  `\&12\catcode `\#12\catcode `\^12\catcode `\_12\catcode `\%12\relax}%
\providecommand \@@startlink[1]{}%
\providecommand \@@endlink[0]{}%
\providecommand \url  [0]{\begingroup\@sanitize@url \@url }%
\providecommand \@url [1]{\endgroup\@href {#1}{\urlprefix }}%
\providecommand \urlprefix  [0]{URL }%
\providecommand \Eprint [0]{\href }%
\@ifxundefined \urlstyle {%
  \providecommand \doi  [0]{\begingroup \@sanitize@url \@doi}%
  \providecommand \@doi [1]{\endgroup \@@startlink {\doibase
  #1}doi:\discretionary {}{}{}#1\@@endlink }%
}{%
  \providecommand \doi  [0]{doi:\discretionary{}{}{}\begingroup
  \urlstyle{rm}\Url }%
}%
\providecommand \doibase [0]{http://dx.doi.org/}%
\providecommand \Doi [0]{\begingroup \@sanitize@url \@Doi }%
\providecommand \@Doi  [1]{\endgroup\@@startlink{\doibase#1}\@@Doi}%
\providecommand \@@Doi [1]{#1\@@endlink}%
\providecommand \selectlanguage [0]{\@gobble}%
\providecommand \bibinfo  [0]{\@secondoftwo}%
\providecommand \bibfield  [0]{\@secondoftwo}%
\providecommand \translation [1]{[#1]}%
\providecommand \BibitemOpen [0]{}%
\providecommand \bibitemStop [0]{}%
\providecommand \bibitemNoStop [0]{.\EOS\space}%
\providecommand \EOS [0]{\spacefactor3000\relax}%
\providecommand \BibitemShut  [1]{\csname bibitem#1\endcsname}%
\bibitem [{\citenamefont {Stauffer}\ and\ \citenamefont
  {Aharony}(1994)}]{Stauffer1994}%
  \BibitemOpen
  \bibfield  {author} {\bibinfo {author} {\bibfnamefont {D.}~\bibnamefont
  {Stauffer}}\ and\ \bibinfo {author} {\bibfnamefont {A.}~\bibnamefont
  {Aharony}},\ }\Doi {10.1103/RevModPhys.63.991} {\emph {\bibinfo {title}
  {Introduction to Percolation Theory}}},\ Vol.~\bibinfo {volume} {1}\
  (\bibinfo {year} {1994})\ ISBN \bibinfo {isbn} {0748402535},\ p.\ \bibinfo
  {pages} {192}\BibitemShut {NoStop}%
\bibitem [{\citenamefont {Grimmett}(1997)}]{Grimmett1997}%
  \BibitemOpen
  \bibfield  {author} {\bibinfo {author} {\bibfnamefont {G.}~\bibnamefont
  {Grimmett}},\ }\Doi {10.1007/BFb0092617} {\emph {\bibinfo {title} {Lectures
  on Probability Theory and Statistics}}},\ edited by\ \bibinfo {editor}
  {\bibfnamefont {P.}~\bibnamefont {Bernard}},\ \bibinfo {series} {Lecture
  Notes in Mathematics}, Vol.\ \bibinfo {volume} {1665}\ (\bibinfo  {publisher}
  {Springer Berlin Heidelberg},\ \bibinfo {year} {1997})\ ISBN \bibinfo {isbn}
  {978-3-540-63190-3}\BibitemShut {NoStop}%
\bibitem [{\citenamefont {Raussendorf}\ and\ \citenamefont
  {Briegel}(2001)}]{mbqc}%
  \BibitemOpen
  \bibfield  {author} {\bibinfo {author} {\bibfnamefont {R.}~\bibnamefont
  {Raussendorf}}\ and\ \bibinfo {author} {\bibfnamefont {H.~J.}\ \bibnamefont
  {Briegel}},\ }\href@noop {} {\bibfield  {journal} {\bibinfo  {journal}
  {Physical Review Letters},\ }\textbf {\bibinfo {volume} {86}},\ \bibinfo
  {pages} {5188} (\bibinfo {year} {2001})}\BibitemShut {NoStop}%
\bibitem [{\citenamefont {Kieling}\ \emph {et~al.}(2007)\citenamefont
  {Kieling}, \citenamefont {Rudolph},\ and\ \citenamefont
  {Eisert}}]{Kieling2007}%
  \BibitemOpen
  \bibfield  {author} {\bibinfo {author} {\bibfnamefont {K.}~\bibnamefont
  {Kieling}}, \bibinfo {author} {\bibfnamefont {T.}~\bibnamefont {Rudolph}}, \
  and\ \bibinfo {author} {\bibfnamefont {J.}~\bibnamefont {Eisert}},\ }\Doi
  {10.1103/PhysRevLett.99.130501} {\bibfield  {journal} {\bibinfo  {journal}
  {Physical Review Letters},\ }\textbf {\bibinfo {volume} {99}},\ \bibinfo
  {pages} {130501} (\bibinfo {year} {2007})}\BibitemShut {NoStop}%
\bibitem [{\citenamefont {Browne}\ and\ \citenamefont
  {Rudolph}(2005)}]{Browne2005}%
  \BibitemOpen
  \bibfield  {author} {\bibinfo {author} {\bibfnamefont {D.~E.}\ \bibnamefont
  {Browne}}\ and\ \bibinfo {author} {\bibfnamefont {T.}~\bibnamefont
  {Rudolph}},\ }\Doi {10.1103/PhysRevLett.95.010501} {\bibfield  {journal}
  {\bibinfo  {journal} {Phys. Rev. Lett.},\ }\textbf {\bibinfo {volume} {95}},\
  \bibinfo {pages} {010501} (\bibinfo {year} {2005})}\BibitemShut {NoStop}%
\bibitem [{\citenamefont {Varnava}\ \emph {et~al.}(2008)\citenamefont
  {Varnava}, \citenamefont {Browne},\ and\ \citenamefont
  {Rudolph}}]{Varnava2008}%
  \BibitemOpen
  \bibfield  {author} {\bibinfo {author} {\bibfnamefont {M.}~\bibnamefont
  {Varnava}}, \bibinfo {author} {\bibfnamefont {D.~E.}\ \bibnamefont {Browne}},
  \ and\ \bibinfo {author} {\bibfnamefont {T.}~\bibnamefont {Rudolph}},\ }\Doi
  {10.1103/PhysRevLett.100.060502} {\bibfield  {journal} {\bibinfo  {journal}
  {Physical Review Letters},\ }\textbf {\bibinfo {volume} {100}},\ \bibinfo
  {pages} {060502} (\bibinfo {year} {2008})}\BibitemShut {NoStop}%
\end{thebibliography}
\end{document}